\documentclass[RNAAS,twocolumn]{aastex63}
\usepackage{amsmath}

\shorttitle{Intrinsic handedness in O1-O4a black-hole mergers}
\shortauthors{Calder\'{o}n Bustillo et al.}
\graphicspath{{./}{figures/}}
\usepackage[figuresright]{rotating}
\usepackage{subfigure}
\usepackage{booktabs}

\begin{document}


\title{Intrinsic handedness in O1-O4a black-hole mergers: \\ probing orbital precession, remnant retention in dense environments and cosmological mirror asymmetry}

\author{Juan Calder\'on~Bustillo}
	\affiliation{Instituto Galego de F\'{i}sica de Altas Enerx\'{i}as, Universidade de
Santiago de Compostela, 15782 Santiago de Compostela, Galicia, Spain}
	\affiliation{Department of Physics, The Chinese University of Hong Kong, Shatin, N.T., Hong Kong}
\author{Adrian del Rio}
\affiliation{Universidad Carlos III de Madrid, Departamento de Matem\'aticas. Avda de la Universidad 30 (edificio Sabatini), 28911 Legan\'es, Spain.}
\author{Nicolas Sanchis-Gual}
\affiliation{Departament d'Astronomia i Astrof\'isica, Universitat de Val\`encia, Avenida Vicent Andr\'es Estell\'es 19, 46100 Burjassot (Val\`encia), Spain}
\author{Koustav Chandra}
    \affiliation{Max Planck Institute for Gravitational Physics (Albert Einstein Institute), Am M\"uhlenberg 1, 14476 Potsdam, Germany}

\begin{abstract}
Precessing binary black-holes generically produce an imbalance of right- and left- handed gravitational waves, reflecting the breaking of mirror symmetry by the merger dynamics. We study this phenomenon using the observer-independent quantity $V_{\rm GW}$, a gravitational analogue of the optical Stokes parameter that quantifies the intrinsic handedness of the emitted radiation. Using 91 LIGO-Virgo-KAGRA black-hole mergers from the O1-O4a observing runs, we find that $92\%$ of the analyzed events favour non-vanishing $V_{\rm GW}$, indicating a predominance of precessing dynamics across the events. Through a recently established relation between $V_{\rm GW}$ and the remnant black hole recoil, we further constrain the retention of merger remnants in dense stellar environments, finding that at most $8\%$ could remain gravitationally bound to globular or nuclear star clusters and subsequently participate in hierarchical merger channels. We finally investigate the cosmological distribution of black-hole merger handedness. The observed $V_{\rm GW}$ distribution is consistent with symmetry under $V_{\rm GW}\rightarrow -V_{\rm GW}$, and yields an average value $\langle V_{\rm GW}\rangle=-1.9^{+6.1}_{-6.6}\times10^{-3}$ ($90\%$ credibility), consistent with the absence of a preferred handedness and with expectations from large-scale statistical isotropy. In particular, the inclusion of O4a events reduces uncertainties in $\langle V_{\rm GW} \rangle$ by $\sim 40\%$ with respect to O1-O3 events. These results establish black-hole merger handedness as a unified probe of orbital precession, remnant recoil, hierarchical formation, and cosmological mirror symmetry.

\end{abstract}



\section{\textbf{Introduction}}
The gravitational-wave (GW) interferometers, Advanced LIGO~\citep{LIGOScientific:2014pky}, Advanced Virgo~\citep{VIRGO:2014yos}, and KAGRA~\citep{KAGRA:2018plz}, have established GW astronomy as a precision probe of compact-object dynamics and the strong-field regime of gravity. Since the first direct detection of a binary black-hole (BBH) merger in 2015~\citep{Abbott:2016blz}, hundreds of additional compact-binary mergers have been reported~\citep{LIGOScientific:2018mvr, abbott2021gwtc2, abbott2021gwtc3, GWTC4-catalogue,GWTC5-Obs}. These have provided unprecedented information on the astrophysical formation channels, merger scenarios and population properties of black holes (BHs) and neutron stars~\citep{Populations_GWTC3, GWTC4_pop,GWTC-5-pop}, the large-scale structure of the Universe \citep{H0_nature_lvk, CaldernBustillo2025_Mirror,GWTC-5-cosmo}, as well as highly non-linear gravitational phenomena including spin precession \cite{Hannam_nature_precession}, higher-order multipoles \cite{GW190412_LVK,Capano2023}, and BH recoil~\citep{Vijay_GWKick, CaldernBustillo2025, kicks_llobera, tousif_kicks}. GW observations have also enabled tests of General Relativity (GR) in the dynamical, strong-curvature regime~\citep{Abbott:2016blz, LIGOScientific:2019fpa, GWTC3-TGR, Isi2019_nohair, Bustillo2021, Siegel2023, TGR_GWTC4_general, TGR_GWTC4_parametrised, remnant, Chandra:2025jfc, Chiaramello:2025bhi, Chandra:2025ipu, LIGOScientific:2025wao}.
As detector sensitivity and waveform modeling continue to improve, current observations are beginning to probe increasingly subtle features of BBH dynamics, including orbital precession and higher-order multipolar structure. 

Recent studies \citep{dRetal20, Sanchis-Gual:2023oqd, Leong2025_Mirror, CaldernBustillo2025_Mirror} have suggested that several seemingly distinct strong-gravity phenomena ---including orbital precession, remnant recoil, GW circular polarization, and parity-odd multipolar asymmetries \citep{Mielke2025,Mielke_2026,kolitsidou2024impact}--- actually encode a common notion of chirality in BBH mergers. This connection {\color{black}arises because} precessing BBHs dynamically break mirror symmetry through the relative orientation and evolution of their spins and orbital angular momentum, leading to intrinsically handed GW emission patterns {\color{black}characterized by an unequal production}  of right- and left- handed gravitational waves. The {\color{black}same underlying asymmetry} is also reflected  in the recoil velocity imparted to the remnant BH by the asymmetric emission of linear momentum in GWs \citep{Gonzalez:2006md,Misner:1974qy}, {\color{black}in particular} through a non-{\color{black}zero} component along its spin {\color{black}direction}. Recent work \citep{CaldernBustillo2025_Mirror} has shown {\color{black}that  these manifestations of chirality can be captured by a single observable, $V_{\rm GW}$ \citep{dRetal20}, a gravitational analogue of the optical Stokes parameter that quantifies the net handedness of the emitted GWs, which has recently been used to probe large-scale mirror symmetry in the observed population of BH mergers.}

\subsection{\textbf{A gravitational Stokes parameter for {\color{black}quantifying} GW polarisation}} Mirror asymmetry in precessing BBHs is reflected in the emission of GWs with net circular polarization, quantified by the observable
\begin{eqnarray}
     && V_{{\rm GW}}= \int_{0}^{\infty} d \omega \,\omega^{3}   \sum_{\ell=2}^{\infty}\sum_{m=-\ell}^{+\ell}\nonumber \\
      &&\left[\bigl|\tilde h^{+}_{\ell m}(\omega)-i \tilde h^{\times}_{\ell m}(\omega)\right|^{2}
-\left|\tilde h^{+}_{\ell m}(\omega)+i \tilde h^{\times}_{\ell m}(\omega)\right|^{2}\bigl] \ .  
 \label{VGW}
\end{eqnarray}

Here, $\tilde h^{+,\times}_{\ell m}(\omega)$ denote the Fourier components of the two GW polarizations appearing in the spin $-2$ spherical-harmonic decomposition of the GW strain, $h_{lm}(t) = h^+_{lm}(t) -i h^{\times}_{lm}(t)$. Originally motivated by geometric considerations in asymptotically flat spacetimes~\citep{dRetal20,dR21}, $V_{\rm GW}$ represents a gravitational analogue of the optical Stokes $V$ parameter, measuring the net circular polarization emitted during the merger. The combinations $\tilde h^{+}_{\ell m}-i \tilde h^{\times}_{\ell m}$ and $\tilde h^{+}_{\ell m}+i \tilde h^{\times}_{\ell m}$ respectively correspond to left- and right-handed circularly polarized GW modes. Under a mirror transformation, these two contributions are exchanged, implying $V_{\rm GW}\to -V_{\rm GW}$. Consequently, mirror-symmetric BBHs satisfy $V_{\rm GW}=0$, {\color{black}while $V_{\rm GW}\neq 0$ indicates intrinsically handed merger dynamics} ~\citep{Sanchis-Gual:2023oqd}. 

\subsection{\textbf{Final BH helicity}} {\color{black}In the binary center-of-mass frame}, 
the remnant BH of a BBH merger carries both a spin $\vec{\chi}_f$ and a recoil velocity $\vec{K}$, naturally defining a pseudoscalar quantity proportional to
\begin{equation}
\begin{aligned}   
     h\sim \vec{\chi}_f \cdot \vec K\, , \label{helicity}
\end{aligned}
\end{equation}
characterizing the handedness -- or helicity -- of the remnant spacetime. This quantity measures the sense of BH rotation relative to its direction of propagation, thus defining {\color{black}another} \textit{intrinsic, observer-independent} notion of handedness for the merger remnant. In contrast to quantities whose apparent orientation depends on the viewing point ---such as the orbital motion of the binary --- the sign of $h$ provides an invariant characterization of the chirality of the system. Systems with $h>0$ and $h<0$ can then be interpreted as right- and left-handed configurations, respectively, depending on whether the resulting final black hole rotates right- or left-handedly along its trajectory, while $h=0$ corresponds to non-chiral configurations. This helicity therefore provides a simple and physically intuitive characterization of the intrinsic chirality generated during the non-linear merger dynamics.\\

\subsection{\textbf{Relation between $V_{\rm GW}$, orbital precession, and BH recoil.}} For aligned-spin binaries, the equatorial symmetry of the system enforces the waveform multipole relation $\tilde h_{\ell,-m}=(-1)^{\ell}\tilde h_{\ell,m}^\ast$, yielding $V_{\rm GW}=0$. Consequently, within the BBH  dynamics, $V_{\rm GW}\neq 0$ requires spin misalignment, which triggers orbital precession. Crucially, while the definitions of circular polarization and helicity (Eqs.~\ref{VGW} and \ref{helicity}) arise from different physical considerations, recent work has shown that {\color{black}the} remnant helicity {\color{black}(\ref{helicity})} and {\color{black} the} GW circular polarization {\color{black}(\ref{VGW})} are in fact linearly correlated, suggesting that both observables encode a common underlying notion of chirality in BBH mergers, imprinted by the handedness of the non-linear spacetime dynamics. 
In particular, \cite{Leong2025_Mirror} found that $V_{\rm GW}$ scales linearly with the projection of the final BH recoil onto its spin axis $(K_a = \vec{K}\cdot\vec{\chi}_f/|\vec{\chi}_f|$) as
\begin{equation}
    V_{\rm GW} = 0.773 \times 10^{-3} K_a [\rm km/s]. 
\label{relation}
\end{equation}
This is consistent with the fact that aligned-spin binaries, for which $V_{\rm GW} = 0$, impart recoils contained in their orbital plane, thus perpendicular to the final spin, yielding $K_a = 0$.

Thus, $V_{\rm GW}$ is a versatile observable with deep implications in both astrophysics and cosmology: it quantifies the net emission of circular polarisation,  probes spin misalignment and orbital precession, {\color{black}characterizes the} handedness of the final BH, and places a lower bound on its recoil velocity. \\

\subsection{\textbf{Astrophysical implications.}} Both orbital precession and gravitational recoil play key roles in deciphering astrophysical BH formation scenarios. Precession is widely considered a smoking gun for dynamical binary formation in dense environments ---such as globular clusters (GCs) and nuclear star clusters (NSCs)--- sourced by their randomly oriented spins. Dense environments are, in turn, expected to enable hierarchical BH formation processes through subsequent mergers \citep{GerosaFishbach, Mandel:2021smh}, which shall explain the observation of both BHs with masses in the pair-instability supernova gap (PISN) \citep{GW190521D, GW231123} -- where no BH formation is expected from stellar collapse \citep{Heger:2002by, Woosley2021} -- and the formation of supermassive BHs \citep{Volonteri2010}. This process is, however, naturally challenged by the recoil velocity of BBH remnants, which can reach thousands of km/s ~\citep{Gonzalez:2006md, PhysRevLett.98.231102, Campanelli_2007, PhysRevD.77.124047, Sperhake:2010uv, Lousto:2011kp} potentially ejecting the remnant BH from host environments such as GCs or NSCs.\\

\subsection{\textbf{Cosmological mirror symmetry.}} GW helicity can serve as a probe of large-scale symmetry.
Under the Cosmological Principle, the Universe is expected to be statistically homogeneous and isotropic on sufficiently large scales, with no preferred handedness in the population of BBH mergers. Consequently, if  {\color{black}right- and left-handed systems form with equal proportions}, the ensemble average of $V_{\rm GW}$ across all mergers should satisfy $\langle V_{\rm GW}\rangle =0$. Non-vanishing values $\langle V_{\rm GW}\rangle$ would therefore signal a large-scale violation of mirror symmetry in the observed BBH population. A first exploratory analysis of this possibility was recently performed in~\cite{CaldernBustillo2025_Mirror} using O1-O3 LVK observations. Importantly, even within the framework of GR ---a fundamentally non-chiral theory--- such studies can probe whether astrophysical formation channels preferentially produce one BBH handedness over the other. Furthermore, because Eq. (\ref{VGW}) is constructed from asymptotic geometric invariants at future null infinity, it provides an intrinsic characterization of the binary spacetime itself, being completely determined by their masses and spins, and fully independent of extrinsic parameters such as the sky location of the source or its orientation relative to the observer \citep{doi:10.1142/S0218271825440043}. This sharply contrasts with several previous approaches to probing large-scale isotropy with gravitational waves~\citep{Salvo_orientations, Essick2023_distributions, Max_directions,Ofek_polarization}, which rely on such observer-dependent quantities.\\

In this work, we study the distribution of $V_{\rm GW}$ across a subset of GW events detected during the O1-O4a LVK observing runs, for which parameter estimation results based on the \texttt{NRSur7dq4} waveform model ~\citep{Varma:2019csw} are publicly available. {\color{black}In particular, we derive} both cosmological and astrophysical implications concerning large-scale mirror symmetry, orbital precession in black-hole mergers and hierarchical black-hole formation in dense environments.

\section{\textbf{Results}}
\subsection{\textbf{Selected events}}
We consider two sets of events detected during the O1-O4a runs of the LVK network, for which parameter inference results are publicly available based on the state-of-the-art waveform model, \texttt{NRSur7dq4}\footnote{Since this model is directly calibrated to numerical relativity simulations obtained by~\cite{SXSCatalog}, naturally incorporating all the corresponding physics, including the asymmetry between positive and negative $m$ modes characteristic of precessing systems. Other phenomenological models that try to reproduce such asymmetry have been recently proposed in \cite{thompson2023phenomxo4a,PhysRevD.109.024061}. The relevance of $m$ asymmetries in GW observations has been examined in e.g. \cite{kolitsidou2024impact,Mielke2025,Mielke_2026,Leong2025_Mirror,Estells2026}}. We consider 47 BBH mergers detected during the O1-O3 period, analysed by~\cite{NRSurCatalog}. The posterior parameter for the masses and spins of these events are available at~\cite{IslamSamples}. To these, we add the 44 O4a that the LVK analysed using the \texttt{NRSur7dq4}, whose samples are publicly available in \cite{GWTC4_data_release}. We obtain posterior distributions for $V_{\rm GW}$ from those on the individual masses and spins through Eq.~\eqref{VGW}, generating the GW modes $h_{\ell m}$ (up to $\ell=4$) using the \texttt{NRSur7dq4} model.\\

\begin{figure}
\begin{center}
\includegraphics[width=0.5\textwidth]{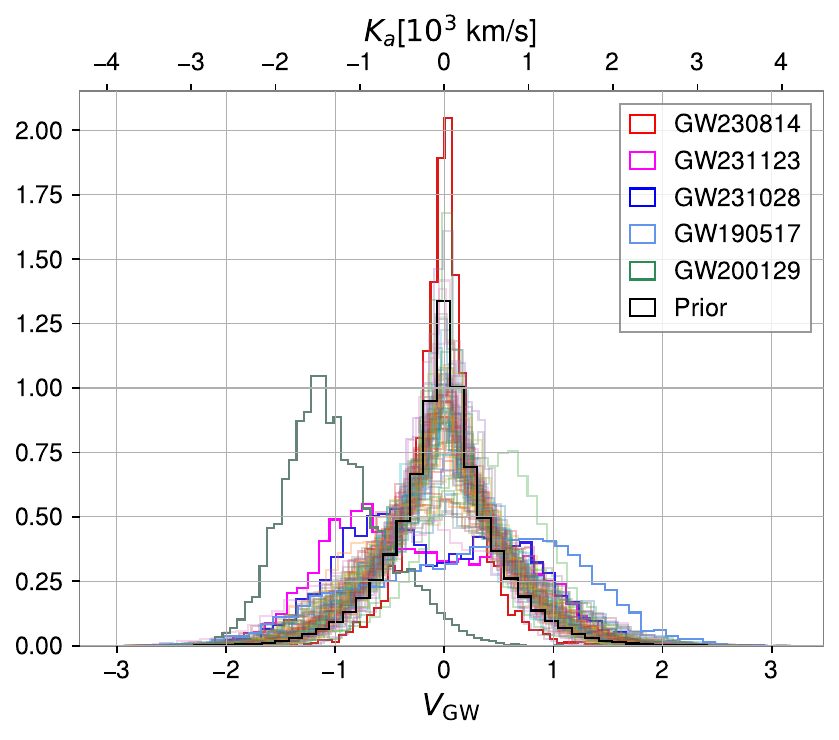}
\caption{\textbf{Posterior probability distribution for the gravitational Stokes parameter (\ref{VGW}) for the 90 events analysed in this work}. The black curve denotes the prior distribution. We highlight the only events displaying the largest evidence for non-zero $V_{\rm GW}$ together with GW230814, which clearly peaks at 0.}
\label{fig:posterior_deltaQJ_all}
\end{center}
\end{figure}

\begin{figure*}
\begin{center}
\includegraphics[width=1\textwidth]{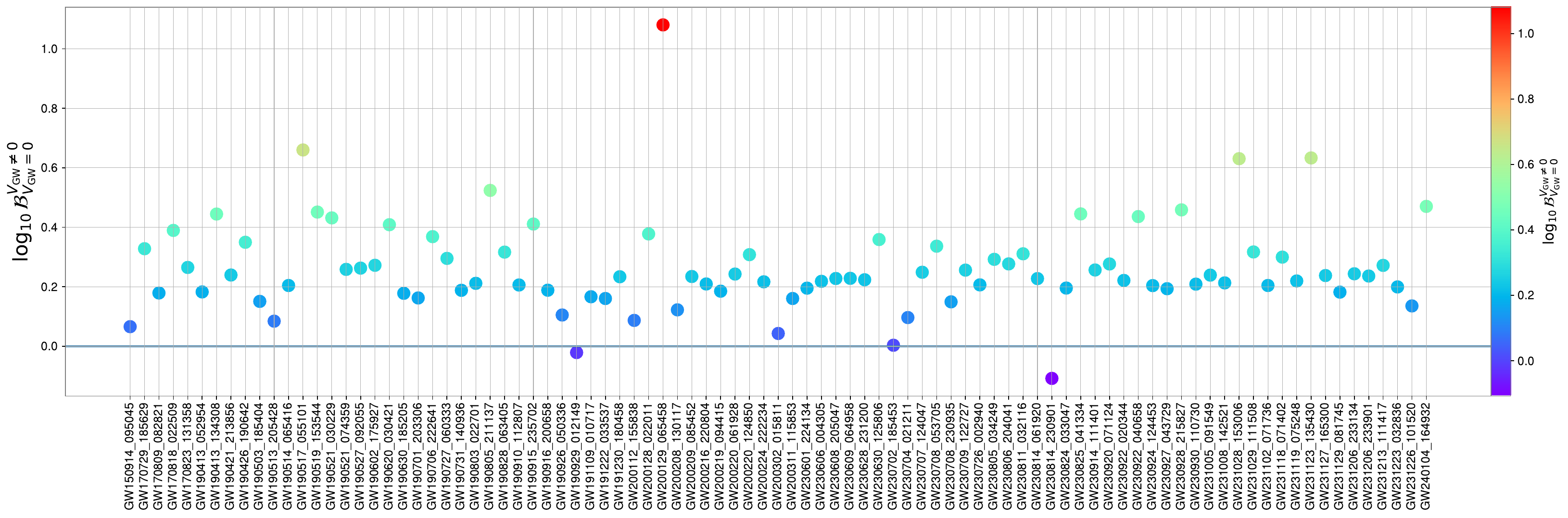}
\caption{\textbf{Evidence for non-zero $V_{\rm GW}$ for each of the analysed events}. For each event, we show the relative Bayes factor for the $V_{\rm GW}\neq 0$ scenario versus the $V_{\rm GW} = 0$ one, based on the publicly available posterior samples.}
\label{fig:bayes}
\end{center}
\end{figure*}

\subsection{\textbf{Individual event results.}}  
Figure~\ref{fig:posterior_deltaQJ_all} shows the individual posterior probability distributions $p(V_{\rm GW}|d)$ for the events we consider. The black curve denotes the prior probability $\pi(V_{\rm GW})$ induced by the priors on the BBH parameters. Among all events, we find that only GW200129, GW190517, GW231028 and GW231123 show significant support for non-zero values, while GW230814 shows the highest peak at $V_{\rm GW} = 0$.
\textcolor{black}{We assess the level to which individual events favour the $V_{\rm GW} = 0$ or $V_{\rm GW} \neq 0$ hypotheses by computing the corresponding Bayesian evidences ratio} ${\cal{B}}^{V_{\rm GW} \neq 0}_{V_{\rm GW} = 0}$, or Bayes Factor. We compute this through Savage-Dickey density ratio \citep{Dickey:1971wlr} evaluated at $V_{\rm GW} = 0$ as ${\cal{B}}^{V_{\rm GW} \neq 0}_{V_{\rm GW} = 0} = \pi(0)/p(0|d)$ (see e.g. \cite{CaldernBustillo2025_Mirror}). The corresponding values are shown in Figure \ref{fig:bayes}. We find that only GW200129 shows strong evidence for $V_{\rm GW} \neq 0$ as reported in \citet{CaldernBustillo2025_Mirror}. This is consistent with the strong evidence for precession found by \cite{Hannam_nature_precession} and the large out-of-plane recoil found by \cite{Vijay_GWKick}. Conversely, GW230814 shows the largest evidence for $V_{\rm GW}=0$, consistently with its sharp posterior peaked at zero (Fig. \ref{fig:posterior_deltaQJ_all}).\\

\subsection{\textbf{Fraction of  events with non-zero $V_{\rm GW}$ and signatures of precession.}} 
\begin{figure}
\begin{center}
\includegraphics[width=0.49\textwidth]{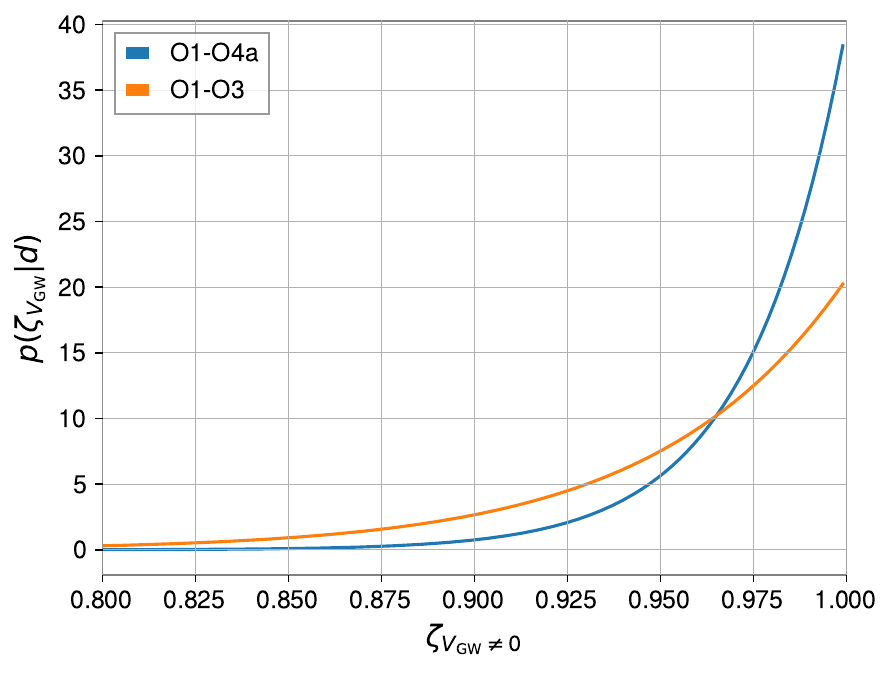}
\caption{\textbf{Fraction $\zeta$ of events of the analysed O1-O4a black-hole mergers with non-null $V_{\rm GW}$}. We show the posterior distributions for O1-O4a events (blue) and O1-O3 events (orange). The prior distribution is uniform in $[0,1]$.}
\label{fig:zetas}
\end{center}
\end{figure}

While only GW200129 displays strong evidence for $V_{\rm GW} \neq 0$, all but two events -- GW230814 and GW190929 -- favour such  hypothesis even if mildly. This suggests that a large fraction $\zeta_{V_{\rm GW}}$ of the events may have $V_{\rm GW}\neq 0$, even if we cannot confidently identify them individually. As done in \citet{CaldernBustillo2025_Mirror}, we compute the likelihood for $\zeta_{V_{\rm GW}}$ as $p(d|\zeta_{V_{\rm GW}}) \propto \prod_{i\in[1,90]} \big{[}{\cal{B}}^{V_{\rm GW}\neq 0}_{V_{\rm GW}=0,i} \zeta_{V_{\rm GW}} + (1-\zeta_{V_{\rm GW}})\big{]}$(~\cite{Callister2022_zeta, Observations_proca,LorenzoMedina2025}. Figure \ref{fig:zetas} shows the corresponding posterior distributions for both O1-O3 events and O1-O4a events. We choose an uniform prior in $\zeta \in [0,1]$. Restricting to O1-O3 we obtain $\zeta_{V_{\rm GW}} = 0.95^{+0.04}_{-0.13}$, indicating that {\color{black}at least} $82\%$ of the events have $V_{\rm GW}\neq0$ \citep{CaldernBustillo2025_Mirror}. Adding the O4a events yields $\zeta_{V_{\rm GW}} = 0.982^{+0.017}_{-0.061}$. This reduces uncertainties by more than $50\%$ and indicates that $92\%$ of the events have $V_{\rm GW}\neq0$. This reinforces the result in \citet{CaldernBustillo2025_Mirror} that the vast majority of the analysed events must display orbital precession, characteristic of dynamical formation in dense environments. This is particularly noteworthy given the observational challenges in robustly detecting precessing BBHs with current matched-filter search techniques \citep{Harry:2016ijz, Bustillo:2016gid, Davies2020, Chandra2020_Nuria}.\\

\subsection{\textbf{Probability of remnant retention in dense environments.}} Eq.~\ref{relation} indicates that $V_{\rm GW}$ places a lower bound on the remnant recoil velocity, $K$. For instance, Fig.~\ref{fig:posterior_deltaQJ_all} shows that the posterior on $K_a$ for GW200129 peaks at 
approximately $1500\, \mathrm{km/s}$. In fact, we obtain $|K_a| = 1736 ^{+782}_{-1120}\, \mathrm{km/s}$. This result is consistent with the kick magnitude $K = 1520 ^{+747}_{-1098}\, \mathrm{km/s}$ inferred by \cite{Vijay_GWKick}.

With this observation, we can place upper limits on the fraction of analyzed remnants that would be retained in environments with a given escape velocity $v_{\rm esc}$ and therefore may contribute to hierarchical BH formation. In particular, GCs and NSCs, which are widely considered candidates to host such formation channels, have escape velocities of about 50 km/s \citep{Merritt:2004xa, Antonini2016} and 1000 km/s \citep{Gerosa2020, Fragione2020}, respectively.\\

\begin{figure}
\begin{center}
\includegraphics[width=0.49\textwidth]{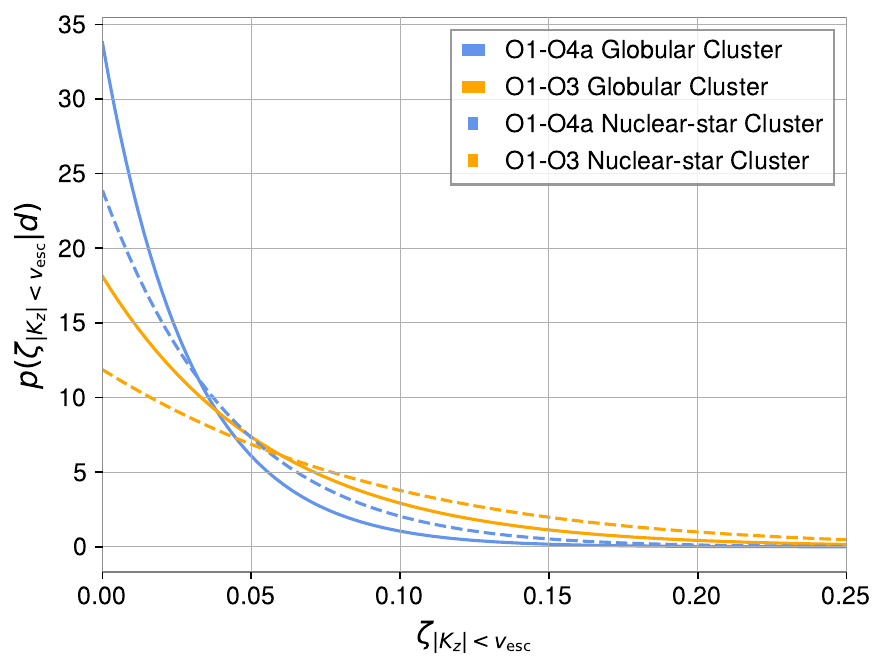}
\caption{\textbf{Fraction of events of the analysed O1-O4a black-hole mergers retained in dense environments}. We show the posterior distributions for O1-O4a fraction $\zeta_{|K_{z}| < v_{\rm esc}}$ of events with a component of the final recoil along the final black-hole spin $|K_{z}|$ is smaller than a given escape velocity $v_{\rm esc}$. We consider $v_{\rm esc}=50$km/s for Globular Clusters (solid) and $v_{\rm esc}=1000$km/s {\color{black}for} Nuclear Star Clusters (dashed). Blue and orange lines correspond to O1-O3 and O1-O4a events, respectively. The prior distribution is uniform in $[0,1]$. Since our calculation only considers the component of the recoil along the spin direction, our results represent upper bounds on the true fraction of retained remnants.}
\label{fig:zetas_kick}
\end{center}
\end{figure}

Using the formalism above, we compute the posterior distribution for the fraction of events $\zeta_{|K_a| < \rm v_{\rm esc}}$ for which $K_a \leq v_{\rm esc}$ for $v_{\rm esc}=50$ km/s and $1000$ km/s, shown in Fig. \ref{fig:zetas_kick} \footnote{We compute individual event Bayes Factors through the formalism in \citet{CaldernBustillo2025_Kick_GW190412}, which generalizes the calculation of the Savage-Dickey ratio to hypotheses defined by extended parameter ranges, in particular by $K_a \in [0,v_{\rm esc}]$. The corresponding values are shown in Figs. \ref{fig:gc} and \ref{fig:nsc} in Appendix II.}. We find that, at most, only a fraction $\zeta_{\rm GC} = 0.020 ^{+0.067}_{-0.018}$ of the remnants could have been retained in GCs, obtaining a slightly larger value $\zeta_{\rm NSC} = 0.029 ^{+0.064}_{-0.018}$ for NSCs. Thus, if the analyzed events occurred in GCs or NSCs, only a small fraction of the remnants would contribute to hierarchical BH formation\footnote{We note that \cite{kicks_llobera} also computed retention probabilities in GCs and NSCs. These, however, are based on posterior probabilities for the recoil of individual events, which are heavily influenced by the prior distribution, peaked at low values. In contrast, we compute actual Bayes Factors.}.\\

\subsection{\textbf{Distribution of $V_{\rm GW}$ across the analysed events}} We study the distribution of $V_{\rm GW}$ over the whole set of analyzed events. To this end, we construct an empirical estimator for the probability density $p_{\rm obs}(\{d_i\}|V_{\rm GW})$ across the whole set of events $\{d_i\}$ based on posterior stacking as:
\begin{equation}
    p_{\rm obs}(\{d_i\}|V_{\rm GW}) \propto \sum_{i} p(d_i|V_{\rm GW}),
\end{equation}
where the likelihood $p(d_i|V_{\rm GW})$ for each individual event is obtained by dividing the posterior distribution by the corresponding prior as
\begin{equation}
     p(d_i|V_{\rm GW}) = \frac{p(V_{\rm GW}|d_i)}{\pi(V_{\rm GW})}.
\end{equation}

\begin{figure}
\begin{center}
\includegraphics[width=0.49\textwidth]{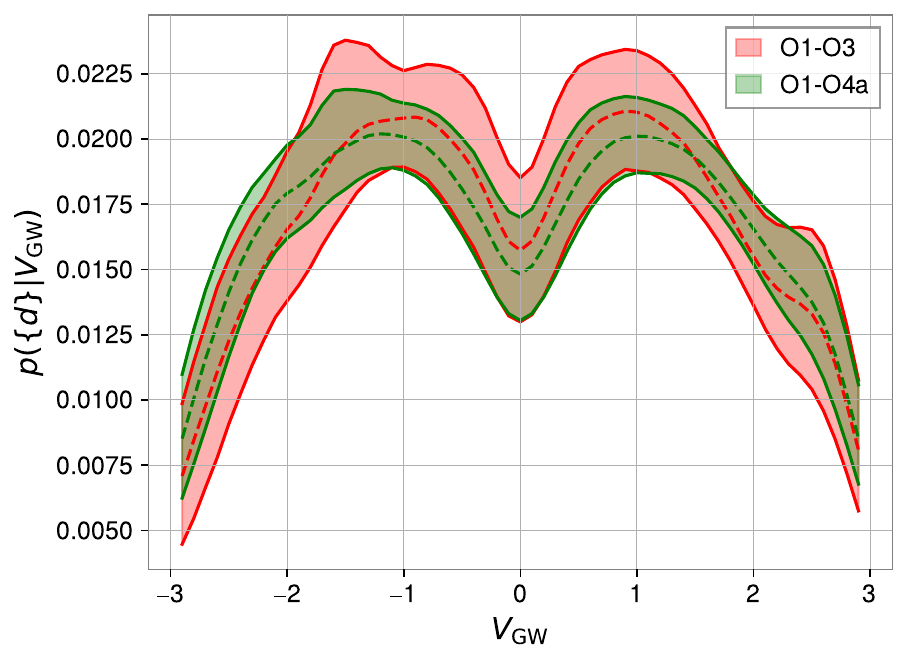}
\caption{\textbf{Distribution of $V_{\rm GW}$ across the analysed black-hole mergers}. We show median values and $90\%$ credible bounds for O1-O4a events (green) and O1-O3 events (red).}
\label{fig:vpop}
\end{center}
\end{figure}

We account for finite observation sample uncertainty by computing $p(\{d_i\}|V_{\rm GW})$ for 1000 random realizations (or bootstraps) of the observed data set, with each realization containing a different random number of copies of each event sampled from a Poisson distribution. This yields 1000 random realizations of the joint likelihood $p_{\rm obs}(\{d_i\}|V_{\rm GW})$. Figure \ref{fig:vpop} shows the corresponding $90\%$ credible intervals for both O1-O3 events and O1-O4a events.\\ 

\subsection{\textbf{Test of mirror symmetry}} We test large-scale mirror symmetry by computing the average value of $V_{\rm GW}$. For each random realisation of $p_{\rm obs}(\{d_i\}|V_{\rm GW})$, we compute 
\begin{equation}
    \langle V_{\rm GW} \rangle = \int_{-\infty}^{\infty} V_{\rm GW} p_{\rm obs}(\{d_i\}|V_{\rm GW}) d V_{\rm GW}.
\end{equation}

\begin{figure}
\begin{center}
\includegraphics[width=0.49\textwidth]{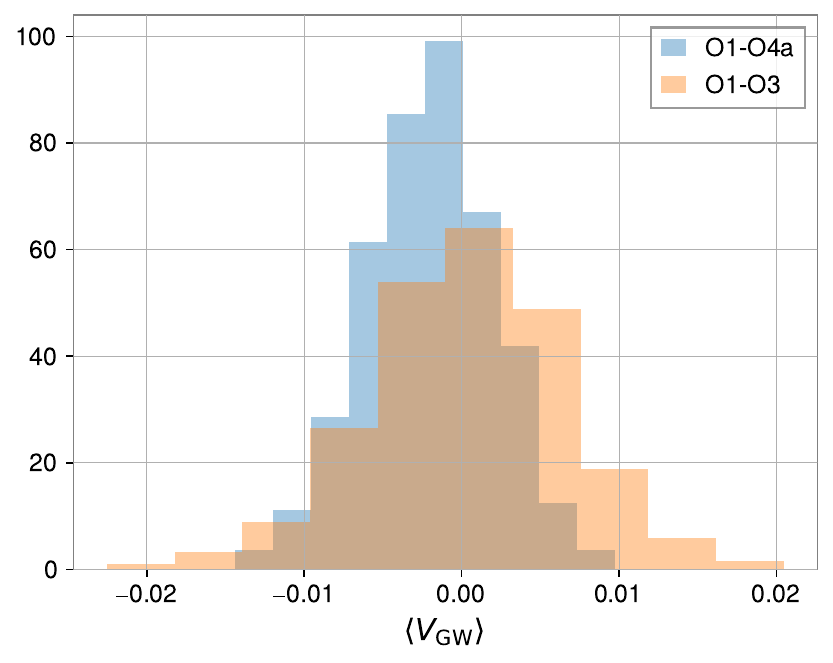}
\caption{\textbf{Mean value of $V_{\rm GW}$ across the analysed O1-O4a black-hole mergers}. We show results for O1-O4a events (blue) and O1-O3 events (orange).}
\label{fig:vmean}
\end{center}
\end{figure}

Figure \ref{fig:vmean} shows the $\langle V_{\rm GW} \rangle$ distribution for both O1-O3 and O1-O4a events. We respectively obtain $\langle V_{\rm GW} \rangle =0.4^{+11.0}_{-10.0} \times 10^{-3}$ and $\langle V_{\rm GW} \rangle =-1.9^{+6.1}_{-6.6} \times 10^{-3}$. Despite the visible shift of the median towards negative values after adding O4a events, both results are fully consistent with $\langle V_{\rm GW} \rangle = 0$, satisfying mirror symmetry and the Cosmological Principle. The addition of O4a events reduces uncertainties by nearly $40\%$. It will be extremely interesting to check if the future inclusion of events from the O4b and O4c observing runs removes or enhances the mentioned median shift, which would then indicate a violation of mirror symmetry.  \\

\subsection{\textbf{Testing $V_{\rm GW} \rightarrow -V_{\rm GW}$ symmetry}}
Visual inspection of Fig.~\ref{fig:vpop} suggests that $V_{\rm GW}$ distributes symmetrically with respect to $V_{\rm GW}=0$. If true, this suggests that astrophysical formation scenarios, which determine the mass ratio and spins of BBHs and thus the value of $V_{\rm GW}$, do not favor a particular helicity. We test this by evaluating the statistical significance of potential deviations from this behavior. To this end, we compare each random realization $p_{\rm obs}^{i}(V_{\rm GW})$ with its reflected counterpart, $p^{i*}_{\rm obs}(V_{\rm GW}) = p^{i}_{\rm obs}(-V_{\rm GW})$. We determine whether the residuals $r_i(V_{\rm GW}) = p_{\rm obs}(V_{\rm GW}) - p^{i*}_{\rm obs}(V_{\rm GW})$ are statistically significant compared to the inherent stochastic uncertainty of the $p_{\rm obs}(V_{\rm GW})$ realizations. 

We estimate the mentioned uncertainty by computing the covariance matrix $\Sigma$ from the ensemble of the 1000 normalized random realizations of the distribution $p(\{d_i\}|V_{\rm GW})$. We then define the test statistic as
\begin{equation}
    \chi^2_{\rm red} = \frac{1}{N \cdot n_{\rm eff}} \sum_{j=1}^{N} \mathbf{r}_j^{\rm T} \Sigma^{-1} \mathbf{r}_j,
\end{equation}
also known as the average Mahalanobis distance.
Here $n_{\rm eff}$ represents the effective degrees of freedom captured via Principal Component Analysis on the set of all random realizations $p^i_{\rm obs}(V_{\rm GW})$. We obtain $\chi^2_{\rm red} \approx 1.1$, indicating that the residuals are consistent with the intrinsic uncertainty of the distribution $p_{\rm obs}(\{d_i\}|V_{\rm GW})$. Thus, we conclude that the data is completely consistent with symmetry around $V_{\rm GW}=0$. We show the original and inverted distributions for $p^i_{\rm obs}(V_{\rm GW})$ in Fig. \ref{fig:symmtest} in Appendix I.\\

\subsection{\textbf{Selection biases and astrophysical population properties}} The above results pertain only to the detected set of events as opposed to the underlying astrophysical population. Obtaining the corresponding distribution $p_{\rm astro}(V_{\rm GW}) \propto p_{\rm obs}(V_{\rm GW}) \times p_{\rm det}(V_{\rm GW})$ requires the estimation of the detection probability $p_{\rm det}(V_{\rm GW})$ of sources with a given value of $V_{\rm GW}$, which we omit.

We have estimated the level to which our conclusions can be extrapolated to the astrophysical population by performing an approximated estimation of $p_{\rm det}(V_{\rm GW})$ (see Appendix III). Although we cannot discard the ``M-shape'' of $p_{\rm obs}(V_{\rm GW})$ in Fig.~\ref{fig:vpop} driven by that of $p_{\rm det}(V_{\rm GW})$, we find that the latter is also symmetric with respect to $V_{\rm GW} = 0$, which drives the following conclusions.

First, the peaks and local minimum of $p_{\rm obs}(V_{\rm{GW}})$ at $V_{\rm GW} \simeq \pm 1$ and $V_{\rm GW} \simeq 0$ may not be population features but selection bias features. Second, the observed $V_{\rm GW} \rightarrow - V_{\rm GW}$ symmetry and the vanishing average $\langle V_{\rm GW} \rangle$ are population properties, since any asymmetry in the astrophysical population would have translated into an asymmetry $p_{\rm obs}(V_{\rm GW})$. Second, the lower limits we have inferred for the fraction of precessing BBHs and BBH remnants ejected from dense environments do not necessarily match those of the astrophysical population. However, we note that any significant suppression of precessing or large recoil systems in nature would necessarily have translated into a corresponding suppression in the observed events. Thus, our results suggest that a large fraction of BBHs within the mass range of the studied events display orbital precession, with their remnants being likely ejected from GCs and NSCs.  \\

\section{\textbf{Discussion}}
$V_{\rm GW}$ is an observer-independent quantity {\color{black}that defines} an \textit{intrinsic notion} of handedness in BBH spacetimes with versatile applications {\color{black}spanning both} cosmology and astrophysics. 

{\color{black}Using 91 BBH mergers from the O1-O4a observing runs, we find strong population-level evidence that the majority of the analysed systems possess non-vanishing values of $V_{\rm GW}$. Since $V_{\rm GW}$ is directly linked to spin misalignment and orbital precession, this result suggests that precessing dynamics are common among the observed BBH population. This picture is broadly consistent with recent studies indicating the presence of dynamically assembled black holes~\citep{Tong2026,Tong_isobel}, including systems populating the PISN gap~\citep{Heger:2002by,Woosley2021}. Through its established connection with the recoil component along the remnant spin, $V_{\rm GW}$ also provides information about the retention of merger remnants in dense stellar environments. We find that the corresponding BH remnants are expected to be efficiently ejected. Specifically, under the assumptions adopted in this work, at most $\sim 8\%$ of the analysed remnants could remain gravitationally bound to GCs or NSCs and subsequently participate in hierarchical merger channels.}

{\color{black}At the same time, our results show no evidence for a preferred handedness} in the observed BBH population, in agreement with expectations of large-scale statistical isotropy. {\color{black}The distribution of $V_{\rm GW}$ is consistent with symmetry under the transformation $V_{\rm GW}\rightarrow -V_{\rm GW}$ and yields an average value compatible with zero.} This is non-trivial:  while Einstein's equations {\color{black}admit}  both right- and left-handed BBH spacetimes, {\color{black}no fundamental principle requires these configurations to occur with equal frequency}. Population asymmetries could arise from certain initial conditions in the early Universe, or from specific astrophysical formation channels that may favor one particular type of handedness, in which case a positive result of our test would not necessarily reflect a fundamental property of gravity. The fact that  $V_{\rm GW}$ is distributed symmetrically around zero, however, suggests that astrophysical formation scenarios do not possess a preferred chirality.

Finally, we note two limiting aspects of our study. First, while we assume that all events are quasi-circular BBHs, orbital precession can be degenerate with eccentricity \citep{Bustillo2021_HeadOn, RomeroShaw2020_ecc_apjl, Gayathri2022_ecc_natastro}.  While this effect, together with certain data-quality issues \citep{Payne2022,Macas2024}, may be behind the $V_{\rm GW}$ value found for GW200129 \citep{Gupte2025,Pompili_ecc}, we have checked that our results are robust against the removal of this event. In addition, our analysis has been restricted to BBHs. Neutron-star mergers may also generate net handedness through post-merger dynamics, even in the absence of strong precession. Such signals could become accessible to next-generation detectors such as the Cosmic Explorer~\citep{CE, CE2}, the Einstein Telescope~\citep{ET1, ET2} or NEMO~\citep{NEMO_Ackley2020}. 

In summary, {\color{black}the O1--O4a catalogue suggests a BBH population that appears largely chiral at the individual-event level, yet statistically symmetric between opposite handednesses. Through the observable $V_{\rm GW}$, this behaviour can be connected within a unified framework to orbital precession, remnant recoil, retention in dense stellar environments, and large-scale mirror symmetry.} As GW catalogs continue to grow and detector sensitivity increases, $V_{\rm GW}$ measurements will place increasingly stringent constraints on {\color{black}the astrophysical origin and cosmological distribution of BBH chirality,} while potentially uncovering new signatures of strong-gravity dynamics.\\

\section{acknowledgments}

We thank Thomas Dent, Xisco Jimenez-Forteza and Gonzalo Morras for useful comments and discussions. JCB is supported by the Ramon y Cajal Fellowship RYC2022-036203-I,
and the research grant PID2020-118635GB-I00 from the Spain-Ministerio de Ciencia e Innovaci\'{o}n. JCB is also supported by the Grant ED431F 2025/04 of the Galician CONSELLERIA DE EDUCACION, CIENCIA, UNIVERSIDADES E FORMACION PROFESIONAL. We also acknowledge support from the European Horizon Europe staff exchange (SE) programme HORIZON-MSCA2021-SE-01 Grant No. NewFunFiCO-101086251.
IGFAE is supported by the Ayuda Maria de Maeztu CEX2023-001318-M funded by MICIU/AEI /10.13039/501100011033. 
ADR acknowledges support through {\it Atraccion de Talento Cesar Nombela} grant No 2023-T1/TEC-29023, funded by Comunidad de Madrid (Spain), and the Spanish Grant  PID2023-149560NB-C21, funded by MCIU/AEI/10.13039/501100011033/FEDER, UE.
NSG acknowledges support from the Spanish Ministry of Science and Innovation via the Ram\'on y Cajal programme (grant RYC2022-037424-I), funded by MCIN/AEI/ 10.13039/501100011033 and by ``ESF Investing in your future”. NSG is further supported by the Spanish Agencia Estatal de Investigaci\'on (Grant PID2021-125485NB-C21) funded by
MCIN/AEI/10.13039/501100011033 and ERDF A way
of making Europe, 
and by the European Horizon Europe staff exchange (SE) programme HORIZON-MSCA2021-SE-01 Grant No. NewFunFiCO-101086251. KC acknowledges the support through NSF grant numbers PHY-2207638, AST-2307147, PHY-2308886, and PHY-2309064.
We acknowledge the use of IUCAA LDG cluster Sarathi for the computational/numerical work. The authors acknowledge computational resources provided by the CIT cluster of the LIGO Laboratory and supported by National Science Foundation Grants PHY-0757058 and PHY0823459; and the support of the NSF CIT cluster for the provision of computational resources for our parameter inference runs. This material is based upon work supported by NSF's LIGO Laboratory which is a major facility fully funded by the National Science Foundation. The analysed LIGO-Virgo data and the corresponding power spectral densities, in their strain versions, are publicly available at the online GW Open Science Center~\cite{SoftwareX,OpenDataArxiv}. This research has made use of data or software obtained from the Gravitational Wave Open Science Center (gwosc.org), a service of LIGO Laboratory, the LIGO Scientific Collaboration, the Virgo Collaboration, and KAGRA. LIGO Laboratory and Advanced LIGO are funded by the United States National Science Foundation (NSF) as well as the Science and Technology Facilities Council (STFC) of the United Kingdom, the Max-Planck-Society (MPS), and the State of Niedersachsen/Germany for support of the construction of Advanced LIGO and construction and operation of the GEO600 detector. Additional support for Advanced LIGO was provided by the Australian Research Council. Virgo is funded, through the European Gravitational Observatory (EGO), by the French Centre National de Recherche Scientifique (CNRS), the Italian Istituto Nazionale di Fisica Nucleare (INFN) and the Dutch Nikhef, with contributions by institutions from Belgium, Germany, Greece, Hungary, Ireland, Japan, Monaco, Poland, Portugal, Spain. KAGRA is supported by Ministry of Education, Culture, Sports, Science and Technology (MEXT), Japan Society for the Promotion of Science (JSPS) in Japan; National Research Foundation (NRF) and Ministry of Science and ICT (MSIT) in Korea; Academia Sinica (AS) and National Science and Technology Council (NSTC) in Taiwan. This manuscript has LIGO DCC number P-2600280. \\

. 

\section{Appendix I: Details on $V_{\rm GW} \rightarrow -V_{\rm GW}$ symmetry} 

\begin{figure}
\begin{center}
\includegraphics[width=0.49\textwidth]{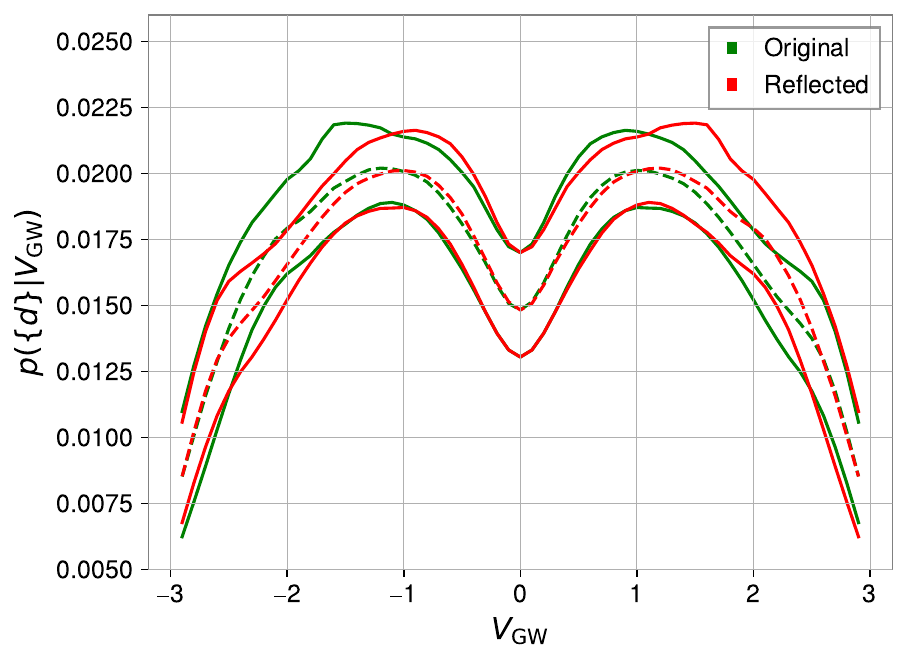}
\caption{\textbf{Symmetry of the distribution for $V_{\rm GW}$ with respect to $V_{\rm GW}=0$}. }
\label{fig:symmtest}
\end{center}
\end{figure}

Figure \ref{fig:symmtest} shows in green the $90\%$ credible bounds for the original distribution for $V_{\rm GW}$ across the studied events for O1-O4a events. In red, we show the same distribution but under the inversion $V_{\rm GW} \rightarrow -V_{\rm GW}$ i.e., reflected with respect to $V_{\rm GW} = 0$. Both distributions show wide agreement, consistently with the conclusions extracted from the $\chi^2$ test described in the main text.

\section{Appendix II: Individual event evidences for retention in dense environments} 

\begin{figure*}
\begin{center}
\includegraphics[width=0.99\textwidth]{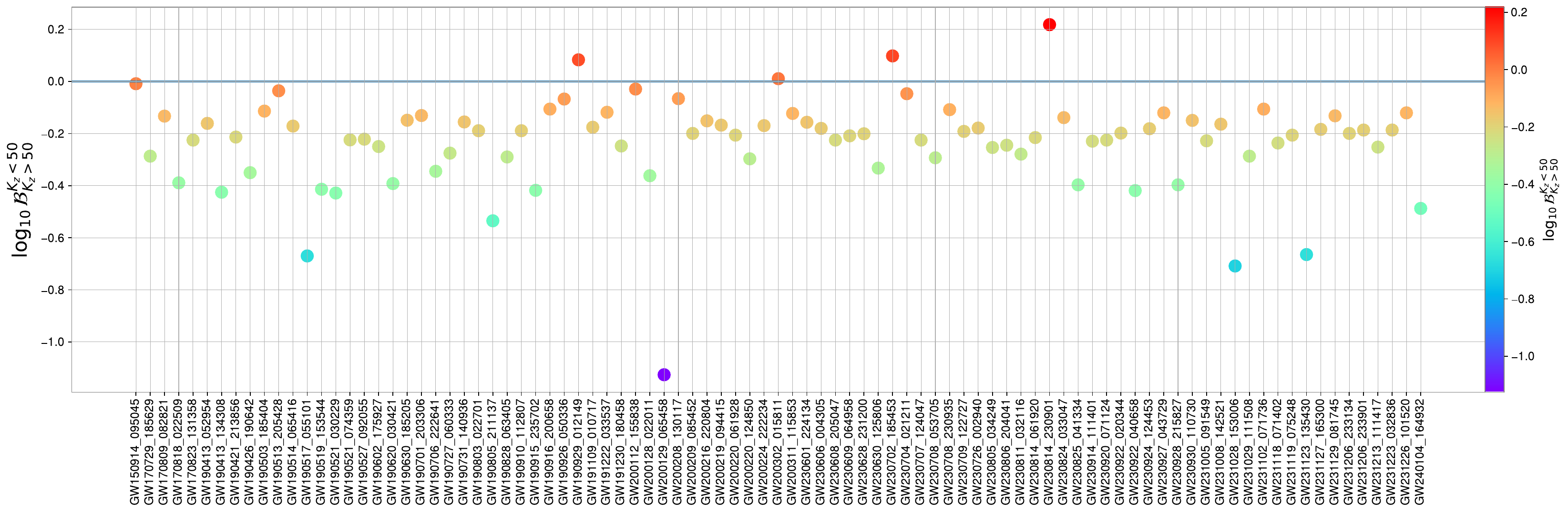}
\caption{\textbf{Event retention in Globular clusters}. We show the individual Bayes Factors for retention versus ejection in Globular Clusters for all the studied events, in base-10 logarithm. }
\label{fig:gc}
\end{center}
\end{figure*}

\begin{figure*}
\begin{center}
\includegraphics[width=0.99\textwidth]{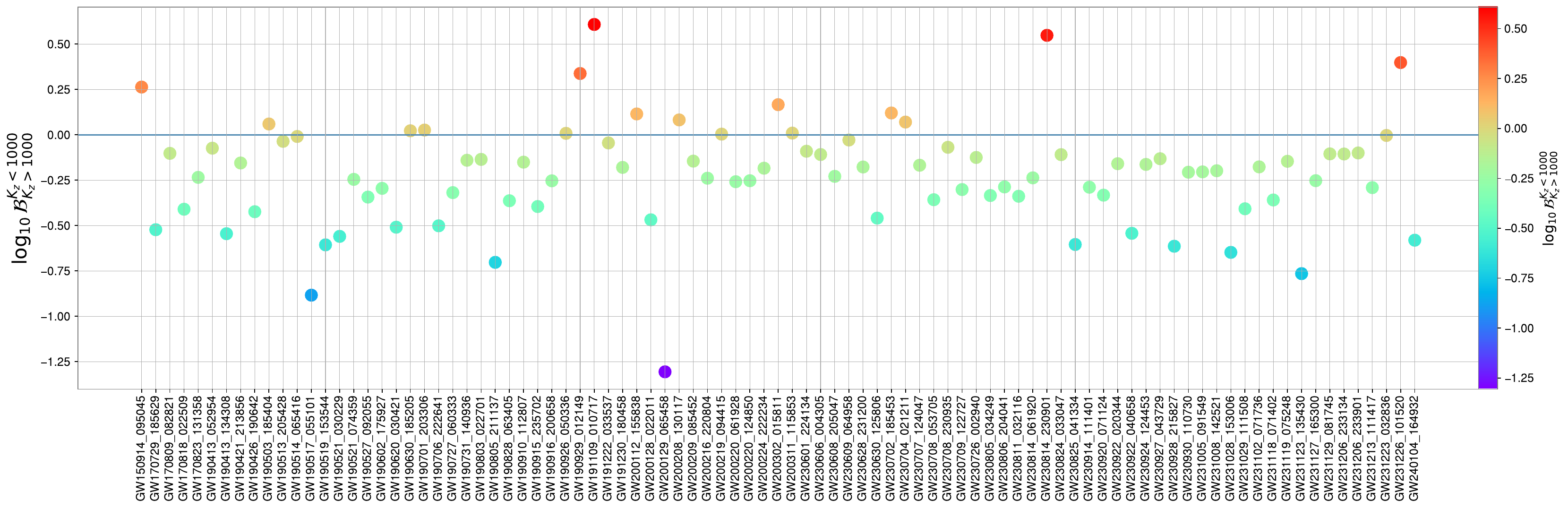}
\caption{\textbf{Event retention in Nuclear Star Clusters}. We show the individual Bayes Factors for retention versus ejection in Nuclear Star Clusters for all the studied events, in base-10 logarithm. }
\label{fig:nsc}
\end{center}
\end{figure*}

Figure \ref{fig:gc} shows the individual event Bayes Factors for models considering ejection and retention in a Globular Cluster, which we characterise by a escape velocity $v_{\rm esc} = 50$ km/s. Fig. \ref{fig:nsc} shows the same for Nuclear Star Clusters, for which we adopt $v_{\rm esc} = 1000$ km/s. In both cases, most event show mild to moderate evidences against retention in these two environments, with GW200129 showing strong evidence for ejection from both cases. GW230814 shows the largest Bayes Factor $\log_{10}\mathcal{B}^{K < 50}_{K>50} \simeq 0.2$ favouring retention in Globular Clusters, which is nevertheless inconclusive. Similarly, GW191109 shows the largest Bayes Factor $\log_{10}\mathcal{B}^{K < 1000}_{K>1000} \simeq 0.5$ favouring retention in Nuclear Star Clusters, which is also inconclusive. In light of these results, it is not surprising that we infer that only a small fraction of the studied events produce remnant black-holes that would be retained in the environments we discuss. 

\section{Appendix III: Selection biases and the astrophysical population.} 
As discussed in the main text, the results we have obtained correspond only to the observed set of events and cannot be, in general, extrapolated to the underlying astrophysical population. The reason is that we have not accounted for the selection bias induced by the detector sensitivity and search algorithms. In general, the probability distribution $p_{\rm astro}(V_{\rm GW})$ of $V_{\rm GW}$ on the true astrophysical population and that on the observed population $p_{\rm obs}(V_{\rm GW})$ are related by $p_{\rm obs}(V_{\rm GW}) = p_{\rm astro}(V_{\rm GW}) \times p_{\rm det}(V_{\rm GW})$, where $p_{\rm det}(V_{\rm GW})$ denotes the probability of detecting an event with a given value of $V_{\rm GW}$. In this Appendix, we perform an approximate estimation of the shape of $p_{\rm det}(V_{\rm GW})$ to then assess the level to which the results we have obtained on the observed events can be extrapolated to the true underlying astrophysical distribution.\\

Formally, an exact estimation of $p_{\rm det}(V_{\rm GW})$ involves the injection a large ensemble of simulated signals into the LVK data and their subsequent recovery through the search pipelines employed by the LVK. Due to the computational cost of such work, we perform here a semi-analytical approximation of $p_{\rm det}$ as follows.\\

We will restrict to the case of search algorithms like \texttt{PyCBC} or \texttt{gstlal}. These algorithms are based on the matched-filter technique, which is the optimal way to search for signals of known morphology. This technique consists on cross-correlating incoming signals $h_{\rm true}$ with large set of template banks composed by pre-computed waveforms $h_t(\theta)$ corresponding to source parameters $\theta$. Under the assumption that the detector noise is Gaussian and stationary, the volumetric sensitivity $V_{\rm sens} \propto p_{\rm det}$ of the search to a given signal $h_{\rm true}(\theta_{\rm true})$ with parameters $\theta_{\rm true}$ is given by $V_{\rm sens} \propto \rho_{\rm MF}^3$. Here, the so-called signal-to-noise ratio (SNR) $\rho_{\rm MF}$ can be expressed as:
\begin{equation}
    \rho_{\rm MF} = \rho_{\rm opt} \times \mathcal{M}.
\end{equation}

Above, $\rho_{\rm opt}$ denotes the maximum SNR that the search can extract from the signal. This will match the actual SNR recovered by the search $\rho_{\rm MF}$ if search template bank contains a template that exactly matches the incoming signal. If not, this will be reduced by a factor $\mathcal{M}\in[0,1]$, known as the match -- or effectualness -- between the incoming signal and the bank template best fitting the signal. This effectualness may be different from one due to a series of reasons. First, searches implement discrete template banks constructed with a minimum match of 0.965 \cite{Nitz_2017,Canton:2014ena,Chandra2022}. In our case, however, we will ignore this fact and assume that banks are infinitely dense. Second, the signal may actually be outside the parameter space covered by the template bank. This is for instance the case of true signals from precessing and eccentric systems, which are currently ignored in LVK searches \cite{Harry:2016ijz}. Finally, even if the signal lays within the parameter space covered by the template bank, waveform models may not reproduce true signals exactly. For instance, in our case, we will assume a template bank formed by waveforms computed by the effective-one-body model \texttt{SEOBNRv4\_ROM} \cite{Bohe:2016gbl}. In addition to orbital precession, this model omits higher-order harmonics and it is ultimately an approximation to the true signal emitted by non-precessing BBHs. All of these effects combined will lead to a finite value of $M$ towards true incoming signals \cite{Varma:2014jxa,Bustillo:2015qty,Bustillo:2016gid,Capano:2013raa}, which we will simulate through the waveform model \texttt{NRSur7dq4}.\\  

Unfortunately, true detector noise is not Gaussian. Instead, it contains spurious noise transients known as glitches that can lead to large SNR values, mimicking the passage of a GW \cite{Canton:2013joa,GravitySpy,Fernandes2023}. Glitches have motivated the development of several signal-glitch discrimination techniques known as vetoes \cite{Allen:2004gu,Nitz:2017lco}, whose output is then combined with the SNR to construct a ranking statistic $\hat\rho$ \cite{Usman:2015kfa,Nitz_2017,2018arXiv181205121M}. If constructed suitably, the ranking statistic will penalise glitches, favouring true GWs. 

Among all the vetoes typically used, we will focus on the so-called Allen $\chi^2$-veto \cite{Allen:2004gu}, which tests the consistency between the expected and observed signal power in different frequency bands. With this, the \texttt{PyCBC} search constructs the quantity $\hat\rho$ as \cite{Nitz2018_pcbclive} 

\[
\hat{\rho} =
\begin{cases}
\rho_{\rm MF}, & \text{for } \chi^2 \leq 1, \\[6pt]
\rho_{\rm MF} \left[ \dfrac{1}{2}\left(1 + \left(\chi^2\right)^3\right) \right]^{-1/6},
& \text{for } \chi^2 > 1 .
\end{cases}
\tag{5}
\]

With this we will approximate the search sensitivity to a given signal by $V_{\rm sens} \propto \hat\rho^3$.\\

We compute $\hat\rho_{\rm sens}$ for $10^{5}$ synthetic signals generated with the \texttt{NRSur7dq4} waveform model, with $V_{\rm GW}$ values uniformly distributed in $[-2,2]$. To this, we first generate $10^7$ random samples from the standard prior parameter distributions imposed by the LVK and then resample uniformly in $V_{\rm GW}$. The restriction to the interval $[-2,2]$ is due to the fact that the number of original samples with values outside this interval is extremely low, as expected from the prior distribution shown in black in Fig. \ref{fig:posterior_deltaQJ_all}. Next, we estimate $V_{\rm sens}$ for each of the resulting $10^{5}$ synthetic signals and compute the average sensitivity $V_{\rm sens}(V_{\rm GW}) = \langle \hat\rho_{\rm sens}^3(\rm V_{\rm GW}) \rangle$ for fixed $V_{\rm GW}$, averaging over all the other signal parameters. Just as in the case of the observed population, we account by finite-sample uncertainty computing $V_{\rm sens}(V_{\rm GW})$ on 1000 independent bootstraps of the $10^{5}$, using \texttt{Poisson}(1) drawing.

\begin{figure}
\begin{center}
\includegraphics[width=0.49\textwidth]{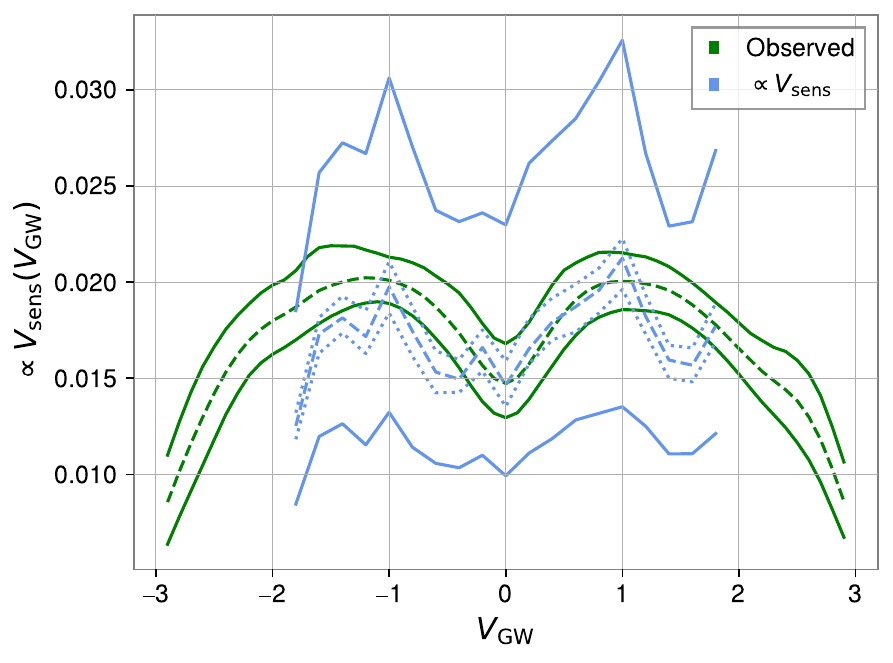}
\caption{\textbf{Observed distribution vs. Selection bias}. The green curves denote the median and $90\%$ credible bounds for the distribution of $V_{\rm GW}$ on the analized 01-o4a events shown in Fig. \ref{fig:vpop}. The blue dashed line and solid contours denote the our median estimate for the marginalised detection probability or sensitive volume $V_{\rm sens}(V_{\rm GW})$, together with the $90\%$ credible bounds. The dotted contour represent the $90\%$ credible bounds on the median.}
\label{fig:pdet}
\end{center}
\end{figure}

Figure~\ref{fig:pdet} shows in green the $90\%$ credible bounds for the distribution of $V_{\rm GW}$ on the analized 01-o4a events shown in Fig. \ref{fig:vpop}. In blue, we show our estimate of the sensitive volume, or detection probability, for each value of $V_{\rm GW }$ averaged over the rest of the source parameters. The dashed line and solid contour denote respectively our median estimate for and the $90\%$ credible bounds. The dotted contour represent the $90\%$ credible bounds on the median. We extract two key conclusions.

First, the shape of $p_{\rm det}$ within the studied interval is broadly consistent with that of the observed population distribution $p_{\rm obs}$. In particular, the observed peaks at $V_{\rm GW} \simeq \pm 1$ and the local minimum at $V_{\rm GW} \simeq 0$ look consistent with the features of $p_{\rm det}$. In particular, our results look compatible with an underlying astrophysical population that is intrinsically flat within $V_{\rm GW} \in [-2,2]$, with the observed structure being a manifestation of the ``sweet spots'' of the detector and search algorithm combination. 

This result may appear counter-intuitive, as current searches primarily target aligned-spin sources, for which $V_{\rm GW} = 0$. While the use of aligned-spin templates leads to a reduction in the template-signal effectualness $\mathcal{M}$, the larger averaged intrinsic loudness of precessing sources causes the optimal SNR $\rho_{\rm opt}$ to rise. Consequently, the combined re-weighted statistic $\hat\rho$ can peak away from $V_{\rm GW} = 0$, effectively favouring the detection of precessing sources despite the lack of dedicated precessing templates in the search banks.

Second, we find that $p_{\rm det}$ is intrinsically symmetric and centred at $V_{\rm GW}=0$. This is a critical result: it implies that the observed symmetry and the vanishing average $\langle V_{\rm GW} \rangle \approx 0$ are not artifacts of selection bias, but rather genuine features of the observed population, as any systematic parity violation in the astrophysical population would have translated into a detectable asymmetry in our sample. This further supports the validity of the Cosmological Principle at these scales.\\

\bibliography{boson_pop,psi4_observation}{}

\begin{thebibliography}{}
\expandafter\ifx\csname natexlab\endcsname\relax\def\natexlab#1{#1}\fi
\providecommand{\url}[1]{\href{#1}{#1}}
\providecommand{\dodoi}[1]{doi:~\href{http://doi.org/#1}{\nolinkurl{#1}}}
\providecommand{\doeprint}[1]{\href{http://ascl.net/#1}{\nolinkurl{http://ascl.net/#1}}}
\providecommand{\doarXiv}[1]{\href{https://arxiv.org/abs/#1}{\nolinkurl{https://arxiv.org/abs/#1}}}

\bibitem[{Aasi {et~al.}(2015)}]{LIGOScientific:2014pky}
Aasi, J., {et~al.} 2015, Class. Quant. Grav., 32, 074001,
  \dodoi{10.1088/0264-9381/32/7/074001}

\bibitem[{Abac {et~al.}(2025{\natexlab{a}})}]{GWTC4-catalogue}
Abac, {et~al.} 2025{\natexlab{a}}, GWTC-4.0: Updating the Gravitational-Wave
  Transient Catalog with Observations from the First Part of the Fourth
  LIGO-Virgo-KAGRA Observing Run.
\newblock \doarXiv{arXiv:2508.18082}

\bibitem[{Abac(2025)}]{GWTC4_pop}
Abac, A.~G. 2025, GWTC-4.0: Population Properties of Merging Compact Binaries.
\newblock \doarXiv{2512.04593}

\bibitem[{Abac {et~al.}(2025{\natexlab{b}})}]{GW231123}
Abac, A.~G., {et~al.} 2025{\natexlab{b}}, The Astrophysical Journal Letters,
  993, L25, \dodoi{10.3847/2041-8213/ae0c9c}

\bibitem[{Abac {et~al.}(2026{\natexlab{a}})}]{TGR_GWTC4_general}
---. 2026{\natexlab{a}}, GWTC-4.0: Tests of General Relativity. I. Overview and
  General Tests.
\newblock \doarXiv{arXiv:2603.19019}

\bibitem[{Abac {et~al.}(2026{\natexlab{b}})}]{TGR_GWTC4_parametrised}
---. 2026{\natexlab{b}}, GWTC-4.0: Tests of General Relativity. II.
  Parameterized Tests.
\newblock \doarXiv{arXiv:2603.19020}

\bibitem[{Abac {et~al.}(2026{\natexlab{c}})}]{remnant}
---. 2026{\natexlab{c}}, GWTC-4.0: Tests of General Relativity. III. Tests of
  the Remnants.
\newblock \doarXiv{arXiv:2603.19020}

\bibitem[{Abac {et~al.}(2026{\natexlab{d}})}]{LIGOScientific:2025wao}
---. 2026{\natexlab{d}}, Phys. Rev. Lett., 136, 041403,
  \dodoi{10.1103/6c61-fm1n}

\bibitem[{Abbott {et~al.}(2020{\natexlab{a}})}]{GW190521D}
Abbott, {et~al.} 2020{\natexlab{a}}, Physical Review Letters, 125,
  \dodoi{10.1103/PhysRevLett.125.101102}

\bibitem[{Abbott {et~al.}(2016)}]{Abbott:2016blz}
Abbott, B.~P., {et~al.} 2016, Phys. Rev. Lett., 116, 061102,
  \dodoi{10.1103/PhysRevLett.116.061102}

\bibitem[{Abbott {et~al.}(2017{\natexlab{a}})}]{H0_nature_lvk}
---. 2017{\natexlab{a}}, Nature, 551, 85, \dodoi{10.1038/nature24471}

\bibitem[{Abbott {et~al.}(2017{\natexlab{b}})}]{CE2}
---. 2017{\natexlab{b}}, Classical and Quantum Gravity, 34, 044001,
  \dodoi{10.1088/1361-6382/aa51f4}

\bibitem[{Abbott {et~al.}(2019{\natexlab{a}})}]{LIGOScientific:2018mvr}
---. 2019{\natexlab{a}}, Phys. Rev. X, 9, 031040,
  \dodoi{10.1103/PhysRevX.9.031040}

\bibitem[{Abbott {et~al.}(2019{\natexlab{b}})}]{LIGOScientific:2019fpa}
---. 2019{\natexlab{b}}, Phys. Rev. D, 100, 104036,
  \dodoi{10.1103/PhysRevD.100.104036}

\bibitem[{Abbott {et~al.}(2020{\natexlab{b}})}]{GW190412_LVK}
Abbott, R., {et~al.} 2020{\natexlab{b}}, Physical Review D, 102,
  \dodoi{10.1103/physrevd.102.043015}

\bibitem[{Abbott {et~al.}(2021{\natexlab{a}})Abbott, Abbott, Abraham, Acernese,
  Ackley, Adams, Adams, Adhikari, Adya, Affeldt, {et~al.}}]{abbott2021gwtc2}
Abbott, R., Abbott, T., Abraham, S., {et~al.} 2021{\natexlab{a}}, Physical
  Review X, 11, 021053

\bibitem[{Abbott {et~al.}(2021{\natexlab{b}})}]{SoftwareX}
Abbott, R., {et~al.} 2021{\natexlab{b}}, {SoftwareX}, 13, 100658,
  \dodoi{10.1016/j.softx.2021.100658}

\bibitem[{Abbott {et~al.}(2023{\natexlab{a}})}]{abbott2021gwtc3}
---. 2023{\natexlab{a}}, Physical Review X, 13,
  \dodoi{10.1103/physrevx.13.041039}

\bibitem[{Abbott {et~al.}(2023{\natexlab{b}})}]{Populations_GWTC3}
---. 2023{\natexlab{b}}, Physical Review X, 13,
  \dodoi{10.1103/physrevx.13.011048}

\bibitem[{Abbott {et~al.}(2025)}]{GWTC3-TGR}
---. 2025, Physical Review D, 112, \dodoi{10.1103/physrevd.112.084080}

\bibitem[{Acernese {et~al.}(2015)}]{VIRGO:2014yos}
Acernese, F., {et~al.} 2015, Class. Quant. Grav., 32, 024001,
  \dodoi{10.1088/0264-9381/32/2/024001}

\bibitem[{Ackley {et~al.}(2020)}]{NEMO_Ackley2020}
Ackley, K., {et~al.} 2020, Publications of the Astronomical Society of
  Australia, 37, \dodoi{10.1017/pasa.2020.39}

\bibitem[{Akutsu {et~al.}(2019)}]{KAGRA:2018plz}
Akutsu, T., {et~al.} 2019, Nature Astron., 3, 35,
  \dodoi{10.1038/s41550-018-0658-y}

\bibitem[{Allen(2005)}]{Allen:2004gu}
Allen, B. 2005, Phys.Rev., D71, 062001, \dodoi{10.1103/PhysRevD.71.062001}

\bibitem[{Antonini \& Rasio(2016)}]{Antonini2016}
Antonini, F., \& Rasio, F.~A. 2016, The Astrophysical Journal, 831, 187,
  \dodoi{10.3847/0004-637x/831/2/187}

\bibitem[{Boh\'e {et~al.}(2017)}]{Bohe:2016gbl}
Boh\'e, A., {et~al.} 2017, Phys. Rev., D95, 044028,
  \dodoi{10.1103/PhysRevD.95.044028}

\bibitem[{Boyle {et~al.}(2019)Boyle, Hemberger, Iozzo, Lovelace, Ossokine,
  Pfeiffer, Scheel, Stein, Woodford, Zimmerman, Afshari, Barkett, Blackman,
  Chatziioannou, Chu, Demos, Deppe, Field, Fischer, Foley, Fong, Garcia,
  Giesler, Hebert, Hinder, Katebi, Khan, Kidder, Kumar, Kuper, Lim, Okounkova,
  Ramirez, Rodriguez, R\"{u}ter, Schmidt, Szilagyi, Teukolsky, Varma, \&
  Walker}]{SXSCatalog}
Boyle, M., Hemberger, D., Iozzo, D. A.~B., {et~al.} 2019, Classical and Quantum
  Gravity, 36, 195006, \dodoi{10.1088/1361-6382/ab34e2}

\bibitem[{Br\"ugmann {et~al.}(2008)Br\"ugmann, Gonz\'alez, Hannam, Husa, \&
  Sperhake}]{PhysRevD.77.124047}
Br\"ugmann, B., Gonz\'alez, J.~A., Hannam, M., Husa, S., \& Sperhake, U. 2008,
  Phys. Rev. D, 77, 124047, \dodoi{10.1103/PhysRevD.77.124047}

\bibitem[{Bustillo {et~al.}(2025)Bustillo, del Rio, Leong, \&
  Sanchis-Gual}]{doi:10.1142/S0218271825440043}
Bustillo, J.~C., del Rio, A., Leong, S. H.~W., \& Sanchis-Gual, N. 2025,
  International Journal of Modern Physics D, 34, 2544004,
  \dodoi{10.1142/S0218271825440043}

\bibitem[{Bustillo {et~al.}(2021{\natexlab{a}})Bustillo, Lasky, \&
  Thrane}]{Bustillo2021}
Bustillo, J.~C., Lasky, P.~D., \& Thrane, E. 2021{\natexlab{a}}, Physical
  Review D, 103, \dodoi{10.1103/physrevd.103.024041}

\bibitem[{Bustillo {et~al.}(2021{\natexlab{b}})Bustillo, Sanchis-Gual,
  Torres-Forné, \& Font}]{Bustillo2021_HeadOn}
Bustillo, J.~C., Sanchis-Gual, N., Torres-Forné, A., \& Font, J.~A.
  2021{\natexlab{b}}, Physical Review Letters, 126,
  \dodoi{10.1103/physrevlett.126.201101}

\bibitem[{Calder{\'o}n~Bustillo {et~al.}(2016)Calder{\'o}n~Bustillo, Husa,
  Sintes, \& P{\"u}rrer}]{Bustillo:2015qty}
Calder{\'o}n~Bustillo, J., Husa, S., Sintes, A.~M., \& P{\"u}rrer, M. 2016,
  Phys. Rev., D93, 084019, \dodoi{10.1103/PhysRevD.93.084019}

\bibitem[{Calder{\'o}n~Bustillo {et~al.}(2017)Calder{\'o}n~Bustillo, Laguna, \&
  Shoemaker}]{Bustillo:2016gid}
Calder{\'o}n~Bustillo, J., Laguna, P., \& Shoemaker, D. 2017, Phys. Rev., D95,
  104038, \dodoi{10.1103/PhysRevD.95.104038}

\bibitem[{Calderón~Bustillo {et~al.}(2025{\natexlab{a}})Calderón~Bustillo,
  del Rio, Sanchis-Gual, Chandra, \& Leong}]{CaldernBustillo2025_Mirror}
Calderón~Bustillo, J., del Rio, A., Sanchis-Gual, N., Chandra, K., \& Leong,
  S.~H. 2025{\natexlab{a}}, Physical Review Letters, 134,
  \dodoi{10.1103/physrevlett.134.031402}

\bibitem[{Calderón~Bustillo {et~al.}(2025{\natexlab{b}})Calderón~Bustillo,
  Leong, \& Chandra}]{CaldernBustillo2025}
Calderón~Bustillo, J., Leong, S. H.~W., \& Chandra, K. 2025{\natexlab{b}},
  Nature Astronomy, 9, 1530–1540, \dodoi{10.1038/s41550-025-02632-5}

\bibitem[{Calderón~Bustillo {et~al.}(2025{\natexlab{c}})Calderón~Bustillo,
  Leong, \& Chandra}]{CaldernBustillo2025_Kick_GW190412}
---. 2025{\natexlab{c}}, Nature Astronomy, 9, 1530–1540,
  \dodoi{10.1038/s41550-025-02632-5}

\bibitem[{Calderón~Bustillo {et~al.}(2023)Calderón~Bustillo, Sanchis-Gual,
  Leong, Chandra, Torres-Forné, Font, Herdeiro, Radu, Wong, \&
  Li}]{Observations_proca}
Calderón~Bustillo, J., Sanchis-Gual, N., Leong, S.~H., {et~al.} 2023, Physical
  Review D, 108, \dodoi{10.1103/physrevd.108.123020}

\bibitem[{Callister {et~al.}(2022)Callister, Miller, Chatziioannou, \&
  Farr}]{Callister2022_zeta}
Callister, T.~A., Miller, S.~J., Chatziioannou, K., \& Farr, W.~M. 2022, The
  Astrophysical Journal Letters, 937, L13, \dodoi{10.3847/2041-8213/ac847e}

\bibitem[{Campanelli {et~al.}(2007{\natexlab{a}})Campanelli, Lousto, Zlochower,
  \& Merritt}]{Campanelli_2007}
Campanelli, M., Lousto, C., Zlochower, Y., \& Merritt, D. 2007{\natexlab{a}},
  The Astrophysical Journal, 659, L5, \dodoi{10.1086/516712}

\bibitem[{Campanelli {et~al.}(2007{\natexlab{b}})Campanelli, Lousto, Zlochower,
  \& Merritt}]{PhysRevLett.98.231102}
Campanelli, M., Lousto, C.~O., Zlochower, Y., \& Merritt, D.
  2007{\natexlab{b}}, Phys. Rev. Lett., 98, 231102,
  \dodoi{10.1103/PhysRevLett.98.231102}

\bibitem[{Capano {et~al.}(2014)Capano, Pan, \& Buonanno}]{Capano:2013raa}
Capano, C., Pan, Y., \& Buonanno, A. 2014, Phys.Rev., D89, 102003,
  \dodoi{10.1103/PhysRevD.89.102003}

\bibitem[{Capano {et~al.}(2023)Capano, Cabero, Westerweck, Abedi, Kastha, Nitz,
  Wang, Nielsen, \& Krishnan}]{Capano2023}
Capano, C.~D., Cabero, M., Westerweck, J., {et~al.} 2023, Physical Review
  Letters, 131, \dodoi{10.1103/physrevlett.131.221402}

\bibitem[{Chandra {et~al.}(2022)Chandra, Bustillo, Pai, \& Harry}]{Chandra2022}
Chandra, K., Bustillo, J.~C., Pai, A., \& Harry, I. 2022, Physical Review D,
  106, \dodoi{10.1103/physrevd.106.123003}

\bibitem[{Chandra \& Calder{\'o}n~Bustillo(2025)}]{Chandra:2025ipu}
Chandra, K., \& Calder{\'o}n~Bustillo, J. 2025.
\newblock \doarXiv{2509.17315}

\bibitem[{Chandra {et~al.}(2025)Chandra, Gamba, \&
  Chiaramello}]{Chandra:2025jfc}
Chandra, K., Gamba, R., \& Chiaramello, D. 2025.
\newblock \doarXiv{2512.04593}

\bibitem[{Chandra {et~al.}(2020)Chandra, Gayathri, Bustillo, \&
  Pai}]{Chandra2020_Nuria}
Chandra, K., Gayathri, V., Bustillo, J.~C., \& Pai, A. 2020, Physical Review D,
  102, \dodoi{10.1103/physrevd.102.044035}

\bibitem[{Chiaramello {et~al.}(2025)Chiaramello, Cibrario, Lange, Chandra,
  Gamba, Bonino, \& Nagar}]{Chiaramello:2025bhi}
Chiaramello, D., Cibrario, N., Lange, J., {et~al.} 2025.
\newblock \doarXiv{2511.19593}

\bibitem[{Collaboration {et~al.}(2023)Collaboration, the Virgo~Collaboration,
  \& the KAGRA~Collaboration}]{OpenDataArxiv}
Collaboration, T. L.~S., the Virgo~Collaboration, \& the KAGRA~Collaboration.
  2023, Open data from the third observing run of LIGO, Virgo, KAGRA and GEO.
\newblock \doarXiv{arXiv:2302.03676}

\bibitem[{Collaboration {et~al.}(2026{\natexlab{a}})Collaboration, the
  Virgo~Collaboration, \& the KAGRA~Collaboration}]{GWTC5-Obs}
---. 2026{\natexlab{a}}, GWTC-5.0: Observations from the Second Part of the
  Fourth LIGO-Virgo-KAGRA Observing Run and Updates to the Gravitational-Wave
  Transient Catalog.
\newblock \doarXiv{arXiv:2605.27225}

\bibitem[{Collaboration {et~al.}(2026{\natexlab{b}})Collaboration, the
  Virgo~Collaboration, \& the KAGRA~Collaboration}]{GWTC-5-pop}
---. 2026{\natexlab{b}}, GWTC-5.0: Population Properties of Merging Compact
  Binaries.
\newblock \doarXiv{arXiv:2605.27226}

\bibitem[{Collaboration {et~al.}(2026{\natexlab{c}})Collaboration, the
  Virgo~Collaboration, \& the KAGRA~Collaboration}]{GWTC-5-cosmo}
---. 2026{\natexlab{c}}, GWTC-5.0: Constraints on the Cosmic Expansion Rate and
  Modified Gravitational-wave Propagation.
\newblock \doarXiv{arXiv:2605.27227}

\bibitem[{Dal~Canton {et~al.}(2014{\natexlab{a}})Dal~Canton, Bhagwat,
  Dhurandhar, \& Lundgren}]{Canton:2013joa}
Dal~Canton, T., Bhagwat, S., Dhurandhar, S., \& Lundgren, A.
  2014{\natexlab{a}}, Class.Quant.Grav., 31, 015016,
  \dodoi{10.1088/0264-9381/31/1/015016}

\bibitem[{Dal~Canton {et~al.}(2014{\natexlab{b}})Dal~Canton, Nitz, Lundgren,
  Nielsen, Brown, {et~al.}}]{Canton:2014ena}
Dal~Canton, T., Nitz, A.~H., Lundgren, A.~P., {et~al.} 2014{\natexlab{b}},
  Phys.Rev., D90, 082004, \dodoi{10.1103/PhysRevD.90.082004}

\bibitem[{Davies {et~al.}(2020)Davies, Dent, Tápai, Harry, McIsaac, \&
  Nitz}]{Davies2020}
Davies, G.~S., Dent, T., Tápai, M., {et~al.} 2020, Physical Review D, 102,
  \dodoi{10.1103/physrevd.102.022004}

\bibitem[{del Rio(2021)}]{dR21}
del Rio, A. 2021, Physical Review D, 104, \dodoi{10.1103/physrevd.104.065012}

\bibitem[{del Rio {et~al.}(2020)del Rio, Sanchis-Gual, Mewes, Agullo, Font, \&
  Navarro-Salas}]{dRetal20}
del Rio, A., Sanchis-Gual, N., Mewes, V., {et~al.} 2020, Physical Review
  Letters, 124, \dodoi{10.1103/physrevlett.124.211301}

\bibitem[{Dickey(1971)}]{Dickey:1971wlr}
Dickey, J.~M. 1971, Annals of Mathematical Statistics, 42, 204.
\newblock \url{http://www.jstor.org/stable/2958475}

\bibitem[{Emtsova \& Birnholtz(2025)}]{Ofek_polarization}
Emtsova, E., \& Birnholtz, O. 2025, Are the Circular Polarizations of Observed
  Gravitational-Waves Even-Handed?
\newblock \doarXiv{arXiv:2508.17103}

\bibitem[{Essick {et~al.}(2023)Essick, Farr, Fishbach, Holz, \&
  Katsavounidis}]{Essick2023_distributions}
Essick, R., Farr, W.~M., Fishbach, M., Holz, D.~E., \& Katsavounidis, E. 2023,
  Physical Review D, 107, \dodoi{10.1103/physrevd.107.043016}

\bibitem[{Estellés {et~al.}(2026)Estellés, Buonanno, Enficiaud, Foo, \&
  Pompili}]{Estells2026}
Estellés, H., Buonanno, A., Enficiaud, R., Foo, C., \& Pompili, L. 2026,
  Physical Review D, 113, \dodoi{10.1103/pjbd-pjxn}

\bibitem[{Fernandes {et~al.}(2023)Fernandes, Vieira, Onofre,
  Calderón~Bustillo, Torres-Forné, \& Font}]{Fernandes2023}
Fernandes, T., Vieira, S., Onofre, A., {et~al.} 2023, Classical and Quantum
  Gravity, 40, 195018, \dodoi{10.1088/1361-6382/acf26c}

\bibitem[{Fragione \& Silk(2020)}]{Fragione2020}
Fragione, G., \& Silk, J. 2020, Monthly Notices of the Royal Astronomical
  Society, 498, 4591–4604, \dodoi{10.1093/mnras/staa2629}

\bibitem[{Gayathri {et~al.}(2022)Gayathri, Healy, Lange, O'Brien,
  Szczepa{\'{n}}czyk, Bartos, Campanelli, Klimenko, Lousto, \&
  O'Shaughnessy}]{Gayathri2022_ecc_natastro}
Gayathri, V., Healy, J., Lange, J., {et~al.} 2022, Nature Astronomy, 6, 344,
  \dodoi{10.1038/s41550-021-01568-w}

\bibitem[{Gerosa \& Fishbach(2021)}]{GerosaFishbach}
Gerosa, D., \& Fishbach, M. 2021, Nature Astronomy, 5, 749–760,
  \dodoi{10.1038/s41550-021-01398-w}

\bibitem[{Gerosa {et~al.}(2020)Gerosa, Vitale, \& Berti}]{Gerosa2020}
Gerosa, D., Vitale, S., \& Berti, E. 2020, Physical Review Letters, 125,
  \dodoi{10.1103/physrevlett.125.101103}

\bibitem[{Ghosh {et~al.}(2024)Ghosh, Kolitsidou, \&
  Hannam}]{PhysRevD.109.024061}
Ghosh, S., Kolitsidou, P., \& Hannam, M. 2024, Phys. Rev. D, 109, 024061,
  \dodoi{10.1103/PhysRevD.109.024061}

\bibitem[{Gonzalez {et~al.}(2007)Gonzalez, Sperhake, Bruegmann, Hannam, \&
  Husa}]{Gonzalez:2006md}
Gonzalez, J.~A., Sperhake, U., Bruegmann, B., Hannam, M., \& Husa, S. 2007,
  Phys. Rev. Lett., 98, 091101, \dodoi{10.1103/PhysRevLett.98.091101}

\bibitem[{Gupte {et~al.}(2025)Gupte, Ramos-Buades, Buonanno, Gair,
  Coleman~Miller, Dax, Green, P\"{u}rrer, Wildberger, Macke, Romero-Shaw, \&
  Sch\"{o}lkopf}]{Gupte2025}
Gupte, N., Ramos-Buades, A., Buonanno, A., {et~al.} 2025, Physical Review D,
  112, \dodoi{10.1103/vpyp-nvfp}

\bibitem[{Hannam {et~al.}(2022)Hannam, Hoy, Thompson, Fairhurst, Raymond,
  Colleoni, Davis, Estell{\'{e}}s, Haster, Helmling-Cornell, Husa, Keitel,
  Massinger, Men{\'{e}}ndez-V{\'{a}}zquez, Mogushi, Ossokine, Payne, Pratten,
  Romero-Shaw, Sadiq, Schmidt, Tenorio, Udall, Veitch, Williams, Yelikar, \&
  Zimmerman}]{Hannam_nature_precession}
Hannam, M., Hoy, C., Thompson, J.~E., {et~al.} 2022, Nature,
  \dodoi{10.1038/s41586-022-05212-z}

\bibitem[{Harry {et~al.}(2016)Harry, Privitera, Boh{\'e}, \&
  Buonanno}]{Harry:2016ijz}
Harry, I., Privitera, S., Boh{\'e}, A., \& Buonanno, A. 2016, Phys. Rev., D94,
  024012, \dodoi{10.1103/PhysRevD.94.024012}

\bibitem[{Heger {et~al.}(2003)Heger, Fryer, Woosley, Langer, \&
  Hartmann}]{Heger:2002by}
Heger, A., Fryer, C.~L., Woosley, S.~E., Langer, N., \& Hartmann, D.~H. 2003,
  Astrophys. J., 591, 288, \dodoi{10.1086/375341}

\bibitem[{Hild {et~al.}(2010)}]{ET2}
Hild, S., {et~al.} 2010, \dodoi{10.1088/0264-9381/28/9/094013}

\bibitem[{Isi {et~al.}(2023)Isi, Farr, \& Varma}]{Max_directions}
Isi, M., Farr, W.~M., \& Varma, V. 2023, The directional isotropy of LIGO-Virgo
  binaries

\bibitem[{Isi {et~al.}(2019)Isi, Giesler, Farr, Scheel, \&
  Teukolsky}]{Isi2019_nohair}
Isi, M., Giesler, M., Farr, W.~M., Scheel, M.~A., \& Teukolsky, S.~A. 2019,
  Physical Review Letters, 123, \dodoi{10.1103/physrevlett.123.111102}

\bibitem[{Islam(2026)}]{tousif_kicks}
Islam, T. 2026, Inference of recoil kicks from binary black hole mergers up to
  GWTC--4 and their astrophysical implications.
\newblock \doarXiv{arXiv:2604.04546}

\bibitem[{Islam {et~al.}(2023)Islam, Vajpeyi, {Feroz Shaik}, Haster, Varma,
  Field, Lange, O'Shaughnessy, \& Smith}]{IslamSamples}
Islam, T., Vajpeyi, A., {Feroz Shaik}, {et~al.} 2023, NRSurCat-1,  Zenodo,
  \dodoi{10.5281/ZENODO.8115310}

\bibitem[{Islam {et~al.}(2025)Islam, Vajpeyi, Shaik, Haster, Varma, Field,
  Lange, O’Shaughnessy, \& Smith}]{NRSurCatalog}
Islam, T., Vajpeyi, A., Shaik, F.~H., {et~al.} 2025, Physical Review D, 112,
  \dodoi{10.1103/48ck-2fff}

\bibitem[{Kolitsidou {et~al.}(2024)Kolitsidou, Thompson, \&
  Hannam}]{kolitsidou2024impact}
Kolitsidou, P., Thompson, J.~E., \& Hannam, M. 2024, Impact of anti-symmetric
  contributions to signal multipoles in the measurement of black-hole spins.
\newblock \doarXiv{2402.00813}

\bibitem[{Leong {et~al.}(2025)Leong, Tomé, Bustillo, del Río, \&
  Sanchis-Gual}]{Leong2025_Mirror}
Leong, S.~H., Tomé, A.~F., Bustillo, J.~C., del Río, A., \& Sanchis-Gual, N.
  2025, Physical Review D, 112, \dodoi{10.1103/1nnp-w5w4}

\bibitem[{{LIGO Scientific Collaboration and Virgo Collaboration and KAGRA
  Collaboration}(2026)}]{GWTC4_data_release}
{LIGO Scientific Collaboration and Virgo Collaboration and KAGRA
  Collaboration}. 2026, GWTC-4.0: Parameter estimation data release,  Zenodo,
  \dodoi{10.5281/ZENODO.16053483}

\bibitem[{Llobera-Querol {et~al.}(2026)Llobera-Querol, Hamilton, Singh,
  Colleoni, Vidal, Askar, Bulik, Olejak, Husa, Xu, \& Valencia}]{kicks_llobera}
Llobera-Querol, J., Hamilton, E., Singh, N., {et~al.} 2026, Remnant recoil and
  host environments of GWTC-4.0 binary black-hole mergers.
\newblock \doarXiv{arXiv:2604.05492}

\bibitem[{Lorenzo-Medina {et~al.}(2025)Lorenzo-Medina, Bustillo, \&
  Leong}]{LorenzoMedina2025}
Lorenzo-Medina, A., Bustillo, J.~C., \& Leong, S.~H. 2025, Physical Review D,
  112, \dodoi{10.1103/lqqf-y7l8}

\bibitem[{Lousto \& Zlochower(2011)}]{Lousto:2011kp}
Lousto, C.~O., \& Zlochower, Y. 2011, Phys. Rev. Lett., 107, 231102,
  \dodoi{10.1103/PhysRevLett.107.231102}

\bibitem[{Macas {et~al.}(2024)Macas, Lundgren, \& Ashton}]{Macas2024}
Macas, R., Lundgren, A., \& Ashton, G. 2024, Physical Review D, 109,
  \dodoi{10.1103/physrevd.109.062006}

\bibitem[{Mandel \& Broekgaarden(2022)}]{Mandel:2021smh}
Mandel, I., \& Broekgaarden, F.~S. 2022, Living Rev. Rel., 25, 1,
  \dodoi{10.1007/s41114-021-00034-3}

\bibitem[{Merritt {et~al.}(2004)Merritt, Milosavljevic, Favata, Hughes, \&
  Holz}]{Merritt:2004xa}
Merritt, D., Milosavljevic, M., Favata, M., Hughes, S.~A., \& Holz, D.~E. 2004,
  Astrophys. J., 607, L9, \dodoi{10.1086/421551}

\bibitem[{Mielke {et~al.}(2026)Mielke, Borchers, \& Ohme}]{Mielke_2026}
Mielke, J., Borchers, A., \& Ohme, F. 2026, Uncovering subdominant multipole
  asymmetries in binary black-hole mergers.
\newblock \doarXiv{arXiv:2602.17343}

\bibitem[{Mielke {et~al.}(2025)Mielke, Ghosh, Borchers, \& Ohme}]{Mielke2025}
Mielke, J., Ghosh, S., Borchers, A., \& Ohme, F. 2025, Physical Review D, 111,
  \dodoi{10.1103/physrevd.111.064009}

\bibitem[{Misner {et~al.}(1974)Misner, Thorne, \& Wheeler}]{Misner:1974qy}
Misner, C.~W., Thorne, K., \& Wheeler, J. 1974

\bibitem[{{Mukherjee} {et~al.}(2018){Mukherjee}, {Caudill},
  {et~al.}}]{2018arXiv181205121M}
{Mukherjee}, D., {Caudill}, S., {et~al.} 2018, arXiv e-prints,
  arXiv:1812.05121.
\newblock \doarXiv{1812.05121}

\bibitem[{Nitz(2017)}]{Nitz:2017lco}
Nitz, A.~H. 2017.
\newblock \doarXiv{1709.08974}

\bibitem[{Nitz {et~al.}(2018)Nitz, Dal~Canton, Davis, \&
  Reyes}]{Nitz2018_pcbclive}
Nitz, A.~H., Dal~Canton, T., Davis, D., \& Reyes, S. 2018, Physical Review D,
  98, \dodoi{10.1103/physrevd.98.024050}

\bibitem[{Nitz {et~al.}(2017)Nitz, Dent, Canton, Fairhurst, \&
  Brown}]{Nitz_2017}
Nitz, A.~H., Dent, T., Canton, T.~D., Fairhurst, S., \& Brown, D.~A. 2017, The
  Astrophysical Journal, 849, 118, \dodoi{10.3847/1538-4357/aa8f50}

\bibitem[{Payne {et~al.}(2022)Payne, Hourihane, Golomb, Udall, Davis, \&
  Chatziioannou}]{Payne2022}
Payne, E., Hourihane, S., Golomb, J., {et~al.} 2022, Physical Review D, 106,
  \dodoi{10.1103/physrevd.106.104017}

\bibitem[{Pompili {et~al.}(2026)Pompili, Gamboa, \& Buonanno}]{Pompili_ecc}
Pompili, L., Gamboa, A., \& Buonanno, A. 2026, Eccentric and unbound compact
  binaries in the LIGO-Virgo-KAGRA catalog: parameter estimation and waveform
  systematics with SEOBNRv6EHM.
\newblock \doarXiv{arXiv:2605.28716}

\bibitem[{Punturo {et~al.}(2010)}]{ET1}
Punturo, M., {et~al.} 2010, Classical and Quantum Gravity, 27, 084007,
  \dodoi{10.1088/0264-9381/27/8/084007}

\bibitem[{Reitze {et~al.}(2019)Reitze, Adhikari, Ballmer, Barish, Barsotti,
  Billingsley, Brown, Chen, Coyne, Eisenstein, Evans, Fritschel, Hall,
  Lazzarini, Lovelace, Read, Sathyaprakash, Shoemaker, Smith, Torrie, Vitale,
  Weiss, Wipf, \& Zucker}]{CE}
Reitze, D., Adhikari, R.~X., Ballmer, S., {et~al.} 2019.
\newblock \doarXiv{arXiv:1907.04833}

\bibitem[{Romero-Shaw {et~al.}(2020)Romero-Shaw, Lasky, Thrane, \&
  Bustillo}]{RomeroShaw2020_ecc_apjl}
Romero-Shaw, I., Lasky, P.~D., Thrane, E., \& Bustillo, J.~C. 2020, The
  Astrophysical Journal, 903, L5, \dodoi{10.3847/2041-8213/abbe26}

\bibitem[{Sanchis-Gual \& del Rio(2023)}]{Sanchis-Gual:2023oqd}
Sanchis-Gual, N., \& del Rio, A. 2023, Phys. Rev. D, 108, 044052,
  \dodoi{10.1103/PhysRevD.108.044052}

\bibitem[{Siegel {et~al.}(2023)Siegel, Isi, \& Farr}]{Siegel2023}
Siegel, H., Isi, M., \& Farr, W.~M. 2023, Physical Review D, 108,
  \dodoi{10.1103/physrevd.108.064008}

\bibitem[{Sperhake {et~al.}(2011)Sperhake, Berti, Cardoso, Pretorius, \&
  Yunes}]{Sperhake:2010uv}
Sperhake, U., Berti, E., Cardoso, V., Pretorius, F., \& Yunes, N. 2011, Phys.
  Rev., D83, 024037, \dodoi{10.1103/PhysRevD.83.024037}

\bibitem[{Thompson {et~al.}(2023)Thompson, Hamilton, London, Ghosh, Kolitsidou,
  Hoy, \& Hannam}]{thompson2023phenomxo4a}
Thompson, J.~E., Hamilton, E., London, L., {et~al.} 2023, PhenomXO4a: a
  phenomenological gravitational-wave model for precessing black-hole binaries
  with higher multipoles and asymmetries.
\newblock \doarXiv{2312.10025}

\bibitem[{Tong {et~al.}(2025)Tong, Callister, Fishbach, Thrane, Antonini,
  Stevenson, Romero-Shaw, \& Dosopoulou}]{Tong_isobel}
Tong, H., Callister, T.~A., Fishbach, M., {et~al.} 2025, A subpopulation of
  low-mass, spinning black holes: signatures of dynamical assembly.
\newblock \doarXiv{arXiv:2511.05316}

\bibitem[{Tong {et~al.}(2026)Tong, Fishbach, Thrane, Mould, Callister, Farah,
  Guttman, Banagiri, Beltran-Martinez, Farr, Galaudage, Godfrey, Heinzel,
  Kalomenopoulos, Miller, \& Vijaykumar}]{Tong2026}
Tong, H., Fishbach, M., Thrane, E., {et~al.} 2026, Nature, 652, 874–877,
  \dodoi{10.1038/s41586-026-10359-0}

\bibitem[{Usman {et~al.}(2016)}]{Usman:2015kfa}
Usman, S.~A., {et~al.} 2016, Class. Quant. Grav., 33, 215004,
  \dodoi{10.1088/0264-9381/33/21/215004}

\bibitem[{Varma {et~al.}(2014)Varma, Ajith, Husa, Bustillo, Hannam,
  {et~al.}}]{Varma:2014jxa}
Varma, V., Ajith, P., Husa, S., {et~al.} 2014, Phys.Rev., D90, 124004,
  \dodoi{10.1103/PhysRevD.90.124004}

\bibitem[{Varma {et~al.}(2019)Varma, Field, Scheel, Blackman, Gerosa, Stein,
  Kidder, \& Pfeiffer}]{Varma:2019csw}
Varma, V., Field, S.~E., Scheel, M.~A., {et~al.} 2019, Phys. Rev. Research., 1,
  033015, \dodoi{10.1103/PhysRevResearch.1.033015}

\bibitem[{Varma {et~al.}(2022)Varma, Biscoveanu, Islam, Shaik, Haster, Isi,
  Farr, Field, \& Vitale}]{Vijay_GWKick}
Varma, V., Biscoveanu, S., Islam, T., {et~al.} 2022, Physical Review Letters,
  128, \dodoi{10.1103/physrevlett.128.191102}

\bibitem[{Vitale {et~al.}(2022)Vitale, Biscoveanu, \&
  Talbot}]{Salvo_orientations}
Vitale, S., Biscoveanu, S., \& Talbot, C. 2022, The orientations of the binary
  black holes in GWTC-3

\bibitem[{Volonteri(2010)}]{Volonteri2010}
Volonteri, M. 2010, The Astronomy and Astrophysics Review, 18, 279,
  \dodoi{10.1007/s00159-010-0029-x}

\bibitem[{Woosley \& Heger(2021)}]{Woosley2021}
Woosley, S.~E., \& Heger, A. 2021, The Astrophysical Journal Letters, 912, L31,
  \dodoi{10.3847/2041-8213/abf2c4}

\bibitem[{Zevin {et~al.}(2017)Zevin, Coughlin, Bahaadini, Besler, Rohani,
  Allen, Cabero, Crowston, Katsaggelos, Larson, Lee, Lintott, Littenberg,
  Lundgren, Østerlund, Smith, Trouille, \& Kalogera}]{GravitySpy}
Zevin, M., Coughlin, S., Bahaadini, S., {et~al.} 2017, Classical and Quantum
  Gravity, 34, 064003, \dodoi{10.1088/1361-6382/aa5cea}

\end{thebibliography}
\bibliographystyle{aasjournal}

\end{document}